\documentclass[11pt]{article}
\pdfoutput=1

\usepackage{jheppub}
\usepackage{amssymb,amsfonts,amsmath,amsthm}
\usepackage{enumerate}
\usepackage{mathrsfs}
\usepackage{bbold}
\usepackage{tikz}
\usetikzlibrary{shapes.geometric, arrows}
\usepackage{comment}
\usepackage{tikz}
\usepackage{hyperref}

\makeatletter
\newcommand{\Vast}{\bBigg@{4.75}}
\makeatother

\newcommand{\be}{\begin{equation}}
\newcommand{\ee}{\end{equation}}
\newcommand{\bea}{\begin{eqnarray}}
\newcommand{\eea}{\end{eqnarray}}

\newcommand{\CA}{\mathcal{A}}

\newcommand{\CC}{\mathcal{C}}
\newcommand{\CD}{\mathcal{D}}

\newcommand{\CG}{\mathcal{G}}

\newcommand{\CI}{\mathcal{I}}

\newcommand{\CK}{\mathcal{K}}
\newcommand{\CL}{\mathcal{L}}
\newcommand{\CN}{\mathcal{N}}
\newcommand{\CM}{\mathcal{M}}
\newcommand{\CO}{\mathcal{O}}
\newcommand{\CP}{\mathcal{P}}
\newcommand{\CQ}{\mathcal{Q}}

\newcommand{\CZ}{\mathcal{Z}}

\newcommand{\lr}{\left (}
\newcommand{\rr}{\right )}
\newcommand{\ls}{\left [}
\newcommand{\rs}{\right ]}

\newcommand\qt\tau

\newcommand{\p}{\partial}

\renewcommand{\tilde}[1]{\widetilde{#1}}
\newcommand{\tr}{\text{tr}}

\makeatletter
\renewcommand{\@seccntformat}[1]{\csname the#1\endcsname.\,\,}
\makeatother
\allowdisplaybreaks

\let \savenumberline \numberline
\def \numberline#1{\savenumberline{#1.}}

\makeatletter
\def\@fpheader{\relax}
\makeatother

\def\bea{\begin{eqnarray}}
\def\eea{\end{eqnarray}}

\usetikzlibrary{shapes,arrows,chains}
\usetikzlibrary{decorations.markings}
\usetikzlibrary{decorations.pathmorphing}
\usetikzlibrary{shapes.multipart}
\tikzset{snake it/.style={decorate, decoration=snake}}

\title{\ \vspace{1.6cm} \\
Strings in Bimetric Spacetimes}
\author{Ziqi Yan}
\emailAdd{ziqi.yan@su.se}
\affiliation{
Nordita, KTH Royal Institute of Technology and Stockholm University\\
Hannes Alfv\'{e}ns v\"{a}g 12, SE-106 91 Stockholm, Sweden}
\abstract{We put forward a two-dimensional nonlinear sigma model that couples (bosonic) matter fields to topological Ho\v{r}ava gravity on a nonrelativistic worldsheet. In the target space, this sigma model describes classical strings propagating in a curved spacetime background, whose geometry is described by two distinct metric fields. We evaluate the renormalization group flows of this sigma model on a flat worldsheet and derive a set of beta-functionals for the bimetric fields. Imposing worldsheet Weyl invariance at the quantum level, we uncover a set of gravitational field equations that dictate the dynamics of the bimetric fields in the target space, where a unique massless spin-two excitation emerges. When the bimetric fields become identical, the sigma model gains an emergent Lorentz symmetry. In this single metric limit, the beta-functionals of the bimetric fields reduce to the Ricci flow equation that arises in bosonic string theory, and the bimetric gravitational field equations give rise to Einstein's gravity.} 

\begin{document}

\maketitle
\vfill\eject

\section{Introduction} \label{sec:intro}

String theory has been serving as a powerful paradigm for addressing fundamental questions in theoretical physics. As a quantum-mechanically consistent framework that admits a massless spin-two excitation, string theory provides a natural arena for testing outstanding ideas about quantum gravity. It has been long known that different superstring theories are corners of a single eleven-dimensional theory of supermembranes dubbed M-theory. However, unlike string theory, which can be defined perturbatively with respect to the genera of Riemann surfaces, there is no simple way of justifying a perturbative expansion in membranes. Another difficulty of defining a quantum theory of membranes comes from the lack of quantization techniques \cite{Horava:1995qa}. It is widely believed that the hypothetical M-theory is inevitably strongly coupled and should be probed by exploring nonperturbative regimes of string theory.

In \cite{Horava:2008ih}, a rather different approach towards a quantum theory of membranes is pioneered, which is designed such that its ground-state wavefunction reproduces the partition function of bosonic string theory. This construction introduces a space and time anisotropy in the worldvolume, and the membranes are described by a three-dimensional nonlinear sigma model (NLSM) at a $z=2$ Lifshitz point, where $z$ denotes the critical dynamical exponent, implying that the worldvolume degrees of freedom enjoy a quadratic dispersion relation and are fundamentally nonrelativistic. The NLSM is coupled to three-dimensional Ho\v{r}ava gravity on the nonrelativistic worldvolume. The worldvolume anisotropy induces a preferred foliation structure by leaves of constant time, which allows one to consistently restrict the sum over three-manifolds in the membrane theory to be a set of more tractable foliated manifolds, bypassing some of the major difficulties in the construction of quantum membranes. 

The membrane theory introduced in \cite{Horava:2008ih}, referred to as \emph{membranes at quantum criticality}, opens up the study of a new spectrum of power-counting renormalizable sigma models coupled to worldvolume Ho\v{r}ava gravity of different dimensions. In the flat worldvolume limit, the worldvolume admits isometries generated by time and space translations, supplemented with the spatial rotations, but there are \emph{no} boost symmetries. This spacetime is nonrelativistic and known as \emph{Aristotelian spacetime} \cite{Penrose:1968ar, Grosvenor:2016gmj}. At renormalization group (RG) fixed points, the sigma models develop various anisotropic scaling invariances, i.e. Lifshitz scaling symmetries, and become of the Lifshitz-type. 
One important step forward is to understand the appropriate target spacetime geometries to which the branes (strings) described by such Lifshitz-type sigma models should be coupled. This requires a thorough analysis of the RG flow structure of these nonrelativistic sigma models, which still remains under-explored. 

In this paper, we focus on the simplest example of the infinite hierarchy of Lifshitz-type sigma models: a two-dimensional sigma model at a $z=1$ Lifshitz point that describes classical strings propagating in spacetime. This is a stringy analogue of the three-dimensional sigma model for membranes at quantum criticality. At the $z=1$ Lifshitz point, the space and time have the same scaling dimension. However, the NLSM is generically nonrelativistic because \emph{no} boost symmetry is imposed \emph{a priori}. On a flat worldsheet, our sigma model is
\be \label{eq:action0}
	S = \frac{1}{4\pi\alpha'} \int_\Sigma dt \, dx \, \Bigl\{ \p_t X^\mu \, \p_t X^\nu \, G_{\mu\nu} (X) - \p_x X^\mu \, \p_x X^\nu \, H_{\mu\nu} (X) \Bigr\}\,,
\ee
where $X^\mu = X^\mu (t, x)$\,, $\mu = 0, 1, \cdots, d-1$ are worldsheet fields that map the worldsheet $\Sigma$ to a $d$-dimensional target space. Without any (neither Lorentzian nor Galilean) boost symmetry on the worldsheet, the background fields $G_{\mu\nu}$ and $H_{\mu\nu}$ are independent symmetric two-tensors. Assuming that these two-tensors are non-degenerate, the geometry of the target space described by $G_{\mu\nu}$ and $H_{\mu\nu}$ has a bimetric feature.\,\footnote{Lifshitz-type sigma models that describe supermembranes propagating in a bimetric spacetime have been discussed in \cite{as}, which we will review later in \S\ref{sec:omqc}.} Although the worldsheet is not boost invariant, the target space is Lorentzian (see \S\ref{eq:smibs}). The action \eqref{eq:action0} defines a unitary and renormalizable quantum field theory. At the RG fixed point where $G_{\mu\nu} = H_{\mu\nu}$\,, the action \eqref{eq:action0} develops an emergent Lorentz symmetry on the worldsheet and underlies standard bosonic string theory, with the target space geometry encoded in a single metric.

For the sigma model \eqref{eq:action0} to describe nonrelativistic strings moving in a bimetric spacetime, we will couple the matter fields $X^\mu$ to appropriate worldsheet gravity that lacks any local boost symmetry, i.e. Ho\v{r}ava gravity in two-dimensions, which is topological and has the notion of a preferred time direction. We will discuss how to couple the sigma model to dilaton fields on a curved worldsheet, at least at the classical level. This forms an essential ingredient towards understanding whether our NLSM is ultimately qualified as a consistent quantum theory of strings that generalize the standard string theory. If there indeed exists a notion of perturbative string theory for the Lifshitz-type NLSM \eqref{eq:action0}, we will have a simple example that goes beyond the framework of relativistic (bosonic) string theory.\,\footnote{Different notions of nonrelativistic strings already exist in the literature. See, e.g., \cite{Gomis:2000bd, Danielsson:2000gi, Batlle:2016iel, Gomis:2016zur, Harmark:2017rpg, Harmark:2018cdl, Bergshoeff:2019pij}. These theories of nonrelativistic strings arise as different corners embedded in relativistic string theory, with various Galilean-type boost symmetries. Along other lines, certain classes of multi-gravity have been shown to arise in relativistic string theory \cite{Kiritsis:2008at}. In contrast, the Lifshitz-type sigma models are beyond relativistic string theory.}  This hypothetical ``bimetric string theory" would define us a bimetric quantum gravity. Alternatively, if there arise any obstacles for defining such bimetric string theory with two distinct spacetime metrics at the quantum level, we would have a rather strong no-go theorem that reinforces the uniqueness of relativistic string theory. 

Regardless whether our sigma model can ultimately be promoted to be a full-fledged quantum theory of strings, the renormalization of the Lifshitz-type sigma model \eqref{eq:action0} already presents a well-defined and challenging problem. 
In this paper, we will mostly focus on the beta-functionals of the background metric fields $G_{\mu\nu}$ and $H_{\mu\nu}$ in \eqref{eq:action0}.\,\footnote{The beta-functionals will be expressed as a Taylor expansion with respect to $G-H$ that we take to be sufficiently small. See \S\ref{sec:pehkc}.} Imposing worldsheet Weyl invariance at the quantum level leads to vanishing beta-functionals, which give rise to a set of field equations that define a novel bimetric gravity. This is analogous to how Einstein's gravity arises in two-dimensional relativistic sigma model on a curved spacetime background. We will analyze the linearized bimetric gravity and reveal that, in the free theory perturbing around flat spacetime, there is one massless spin-two gauge field, together with other tensorial degrees of freedom that satisfy the Klein-Gordon equation. This RG calculation not only provides us with an opportunity for probing exotic bimetric geometries, but also constitutes essential first steps towards a vast landscape of generically nonrelativistic theories of strings and membranes,\,\footnote{In the particle case, however, the sigma model is defined on a worldline and identical to the relativistic case, with a unique metric encoding the target space geometry.}  where relativistic string theory only emerges at a corner with Lorentz symmetry. 

The paper is organized as follows. In \S\ref{sec:csbg}, we define our sigma model coupled to two-dimensional Ho\v{r}ava gravity on the worldsheet, and show that this sigma model describes strings moving in a bimetric spacetime when a time-reversal symmetry is imposed. In \S\ref{sec:rbsm}, we use the heat kernel method to compute the one-loop beta-functions of the bimetric sigma model on a flat worldsheet, in terms of a Taylor expansion with respect to the difference between the metric fields. This result is summarized in \eqref{eq:betaGH}. In \S\ref{sec:omqc}, we discuss generalizations of the string sigma model to classical theories of membranes at quantum criticality. We conclude the paper in \S\ref{sec:concl}. In Appendix \ref{app:chkc}, we present the full result of the heat kernel coefficient that is relevant to the beta-function calculation.

\section{Classical Strings in a Bimetric Geometry} \label{sec:csbg}

In this section, we construct a NLSM that maps a two-dimensional nonrelativistic worldsheet $\Sigma$ to a $d$-dimensional spacetime manifold $\CM$ equipped with two metric fields. We first define the desired nonrelativistic symmetries of the worldsheet, generically excluding any (Lorentzian nor Galilean) boost transformations. The dynamics of the worldsheet geometry is described by two-dimensional Ho\v{r}ava gravity at a $z=1$ Lifshitz point, which is topological. We will couple scalar fields to this two-dimensional nonrelativistic gravity and build up a sigma model that describes classical strings moving in a bimetric geometry. 

\subsection{Elements on \texorpdfstring{Ho\v{r}ava}{Horava} gravity} \label{sec:ehg}

We start with a brief review of Ho\v{r}ava gravity following \cite{Horava:2008ih, Horava:2009uw}, which we will use later to describe the worldsheet geometry. Ho\v{r}ava gravity lives on a $(D+1)$-dimension spacetime manifold $\Sigma$ equipped with a foliation by leaves of codimension one, which are slices of constant time. We will use the coordinates $\bigl(t, \mathbf{x} \! = \! (x^i, i = 1, \cdots\!, D) \bigr)$ that are adapted to the foliation structure. The dynamics of this foliated geometry is described by the ADM formalism variables, originally introduced in the Hamiltonian formulation of general relativity, where the relativistic spacetime metric is decomposed into the lapse function $N$\,, the shift vector $N_i$\,, and the spatial metric $\gamma_{ij}$ \cite{Arnowitt:1959ah}. In Ho\v{r}ava gravity, the ADM variables are used to define the time length element
$ds_{\text{T}}^2 = N^2 \, dt^2$\,,
and the space length element 
$ds_{\text{L}}^2 = \gamma_{ij} \lr N^i dt + d x^i \rr \lr N^j dt + d x^j \rr$,	
which are \emph{a priori} unrelated \cite{Frenkel:2020djn}.
The time (space) length is measured in the time (space) unit $T$ ($L$)\,, with
\be
	\text{dim} (ds^{}_\text{T}) = \text{dim}(t) = T\,, 
		\qquad
	\text{dim} (ds^{}_\text{L}) = \text{dim} (\mathbf{x}) = L\,. 
\ee
It follows that the classical dimensions of the ADM variables are
\be
	\text{dim} (N) = 1\,,
		\qquad
	\text{dim} (N_i) = L/T\,,
		\qquad
	\text{dim} (\gamma_{ij}) = 1\,. 
\ee
At RG fixed points, the classical theory develops anisotropic scale invariance with the dynamical critical exponent $z$\,, such that $T \sim L^z$, which leads to the scaling dimensions for the spacetime coordinates,
\be
	[t] = -1\,,
		\qquad
	[\mathbf{x}] = - z^{-1}\,,
\ee
and for the ADM variables,
\be
	[N] = 0\,,
		\qquad
	[N_i] = 1 - z^{-1}\,,
		\qquad
	[\gamma_{ij}] = 0\,.
\ee
In this convention, energy is of scaling dimension one. 

The gauge symmetries in Ho\v{r}ava gravity are diffeomorphisms that preserve the foliation structure. These foliation-preserving diffeomorphisms act on the spacetime coordinates as
\be \label{eq:fp}
	t \rightarrow t' (t)\,,
		\qquad
	\mathbf{x} \rightarrow \mathbf{x'} (t\,, \mathbf{x})\,.
\ee
Infinitesimally, we write $\delta t = \zeta(t)$ and $\delta x^i = \xi^i (t, \mathbf{x})$\,. 
The ADM variables transform under the infinitesimal foliation-preserving diffeomorphisms as
\begin{subequations} \label{eq:fpd}
\begin{align}
	\delta N & = \p^{}_t \! \lr \zeta \, N \rr + \xi^i \, \p_{i} N\,, \\[2pt]
	\delta N_i & = \p^{}_t \! \lr \zeta \, N_i \rr + \xi^j \, \CD_{\!j} N_i + N_j \, \CD_{i} \, \xi^j + \gamma_{ij} \, \p_t \, \xi^j, \\[2pt]
	\delta \gamma_{ij} & = \zeta \, \p^{}_t \gamma_{ij} + \CD_{i} \, \xi_j + \CD_{\!j} \, \xi_i\,.
\end{align}
\end{subequations}
We have defined the spatial covariant derivative $\CD_{i}$\,, which is defined with respect to the spatial metric $\gamma_{ij}$\,, satisfying the compatible condition $\CD^{}_{\!i} \gamma^{}_{jk} = 0$\,. Quantities that transform covariantly with respect to the foliation-preserving diffeomorphims include the extrinsic curvature,
\be
	K_{ij} = \frac{1}{2 \, N} \lr \p_t {\gamma}_{ij} - \CD_{i} N_j - \CD_{j} N_i \rr,
\ee
the spatial Riemann curvature tensor $R^k{}_{\ell ij}$\,, the acceleration $a_i = - N^{-1} \, \CD_{i} N$, and the covariant time derivative $d_n$\,. Here, $d_n$ is defined to be the Lie derivative with respect to the vector field 
\be \label{eq:nvf}
	n = \frac{1}{N} \lr \p_t - N^i \, \p_i \rr,
\ee
and projected onto the spatial slice. 
We will also impose the following time-reversal symmetry:
\be \label{eq:trs}
	t \rightarrow - t\,,
		\qquad
	N \rightarrow N,
		\qquad
	N_i \rightarrow - N_i\,,
		\qquad
	\gamma_{ij} \rightarrow \gamma_{ij}\,,
\ee
which forbids terms that are odd in time derivatives. In addition, we impose the anisotropic Weyl invariance \cite{Horava:2008ih},
\begin{align} \label{eq:aWi}
	N \rightarrow e^{z \, W(t,\, \mathbf{x})} N, 
		\qquad%
	N_i \rightarrow e^{2 \, W(t,\, \mathbf{x})} N_i\,,
		\qquad%
	\gamma_{ij} \rightarrow e^{2 \, W(t,\, \mathbf{x})} \, \gamma_{ij}\,.
\end{align}
We will soon construct the NLSM that couples scalar fields to two-dimensional Ho\v{r}ava gravity. Requiring that the anisotropic Weyl invariance \eqref{eq:aWi} hold classically as well as at quantum level will give rise to a set of consistency equations that dictate the dynamics of the target space geometry.

In flat limit, we have the ground state solution $N = 1$\,, $N_i = 0$\,, and $\gamma_{ij} = \delta_{ij}$\,. This ground state solution possesses the isometries
\be \label{eq:iso}
	\delta t = \zeta\,,
		\qquad
	\delta x^i = \xi^i + \omega^i{}_j \, \xi^j\,,
\ee
where $\zeta$ and $\xi^i$ parametrize translations in temporal and spatial directions, respectively, and $\omega^i{}_j$ parametrizes spatial rotations. Such spacetime without any (Lorentzian or Galilean) boost symmetry is Aristotelian spacetime \cite{Penrose:1968ar, Grosvenor:2016gmj}. In the special case when $z=1$\,, however, the isometry group of the ground state solution is enlarged to the Poincar\'{e} group that incorporates the Lorentzian boosts. 

\subsection{Dynamics of worldsheet geometry} \label{sec:dwg}

We define the worldsheet to be a two-dimensional manifold $\Sigma$\,, equipped with a codimension\,\!-one foliation structure. The coordinates on the worldsheet are $(t, x \equiv x^1)$\,. We also require the foliation-preserving diffeomorphisms \eqref{eq:fp} on the worldsheet, 
\be \label{eq:fpx}
	t \rightarrow t' (t)\,,
		\qquad
	x \rightarrow x' (t\,, x)\,.
\ee
Infinitesimally, we write
\be \label{eq:fpxinf}
	\delta t = \zeta (t)\,,
		\qquad
	\delta x = \xi (t\,, x)\,.
\ee
Note that we have suppressed a raised ``$x$" index in $\xi$. Under the infinitesimal foliation-preserving diffeomorphisms, from \eqref{eq:fpd} we find that the ADM variables transform as
\begin{subequations} \label{eq:fpdx}
\begin{align}
	\delta N & = \p^{}_t \! \lr \zeta \, N \rr + \xi \, \p_{x} N\,, \\[2pt]
	\delta \chi & = \p^{}_t \! \lr \zeta \, \chi \rr + \p_{x} \bigl(  \xi \, \chi \bigr) + \gamma \, \p_t \xi\,, \label{eq:fpdNx} \\[2pt]
	\delta \gamma & = \zeta \, \p^{}_t \gamma + 2 \, \sqrt{\gamma} \, \p_x \bigl( \sqrt{\gamma} \, \xi \bigr) \,.
\end{align}
\end{subequations}
We defined the shift function $\chi \equiv N_x$ and the spatial metric $\gamma \equiv \gamma_{xx}$\,.
The dynamics of the worldsheet geometry is then described by two-dimensional Ho\v{r}ava gravity that we have briefed in \S\ref{sec:ehg} with $D = z = 1$\,, such that the theory is at its lower critical dimension. 

Imposing the time-reversal symmetry
\be \label{eq:trsx}
	t \rightarrow - t\,,
		\qquad
	N \rightarrow N,
		\qquad
	\chi \rightarrow - \chi\,,
		\qquad
	\gamma \rightarrow \gamma\,,
\ee
we classify independent terms that transform as scalars under the foliation-preserving diffeomorphisms in \eqref{eq:fpdx} and are of scaling dimension two:
\be \label{eq:Ka}
	K^2\,, 
		\qquad
	d_n K\,,
		\qquad
	\gamma^{-1} \, a^2\,,
		\qquad
	\gamma^{-1} \, \CD_x \, a = \frac{1}{\sqrt{\gamma}} \, \p_x \! \lr \frac{a}{\sqrt{\gamma}} \rr,
\ee
where
\begin{subequations}
\begin{align}
	K & = \frac{1}{N \sqrt{\gamma}} \ls \p_t \sqrt{\gamma} - \p_{x} \! \lr \frac{\chi}{\sqrt{\gamma}} \rr \rs \!, \\[2pt]
	d_n K & = \frac{1}{N} \lr \! \p_t - \frac{\chi}{\gamma} \, \p_x \! \rr \! K, \\[2pt]
	a & \equiv a_x = - \frac{1}{N} \, \p_x N.
\end{align}
\end{subequations}
The anisotropic Weyl transformation \eqref{eq:aWi} reduces to
\begin{align} \label{eq:Weylz1}
	N \rightarrow e^{W(t,\, \mathbf{x})} N,
		\qquad
	\chi \rightarrow e^{2 \, W(t,\, \mathbf{x})} \chi\,,
		\qquad
	\gamma \rightarrow e^{2 \, W(t,\, \mathbf{x})} \, \gamma\,.
\end{align}
There are two independent linear combinations among the ingredients in \eqref{eq:Ka} that are invariant under the Weyl transformation \eqref{eq:Weylz1}, namely \cite{Arav:2014goa},
\begin{align}
	K^2 + d_n K\,,
		\qquad
	\gamma^{-1} \bigl( a^2 - \CD_{x} \, a \bigr)\,.
\end{align}
The two-dimensional Weyl-invariant Ho\v{r}ava gravity is described by the following action:
\be \label{eq:Sg}
	S_\text{gr.} = \frac{1}{\kappa^2} \int_\Sigma dt \, dx \, N \sqrt{\gamma} \, \Bigl[ (K^2 + d_n K) - \alpha \, \gamma^{-1} (a^2 - \CD_{x} \, a) \Bigr]\,,
\ee
with two independent coupling constants $\kappa$ and $\alpha$\,. For simplicity, we have taken $\Sigma$ to be compact and boundaryless, and thus \eqref{eq:Sg} does not include any boundary curvature terms. This theory is purely topological, which is manifested by rewriting $S_\text{gr.}$ as
\be
	S_\text{gr.} = \frac{1}{\kappa^2} \int_\Sigma dt \, dx \, \Bigl\{ \p^{}_t \bigl( \gamma^{1/2} \, K \bigr) - \p_x \bigl[ \gamma^{-1/2} \lr \chi K - \alpha \, N a \rr \bigr] \Bigr\}\,,
\ee
where the integrand only contains total derivative terms. 

In the special case when $\alpha = 1$\,, we find 
\be
	S_\text{gr.} = \frac{1}{2 \kappa^2} \int_\Sigma dt \, dx \, \sqrt{- h} \, R(h)\,, 
\ee
where $R(h)$ is the Ricci scalar defined with respect to a worldsheet metric $h_{\alpha\beta}$\,, with
\be \label{eq:hab}
	h_{\alpha\beta} = 
	\begin{pmatrix}
		- N^2 + \gamma^{-1} \, \chi^2 & \,\, \chi \\
		\chi & \,\, \gamma
	\end{pmatrix}
\ee
and $h = \det h_{\alpha\beta}$\,. This is simply the two-dimensional Einstein-Hilbert action, which is conformally invariant. At this relativistic fixed point, the foliation-preserving diffeomorphisms are extended to the full relativistic diffeomorphisms.

\subsection{Sigma models in a bimetric spacetime} \label{eq:smibs}

We are ready to couple scalar fields to the worldsheet geometry presented in \S\ref{sec:dwg}, and construct the sigma model that describes classical strings propagating in a bimetric geometry. 

Define the worldsheet scalar fields $X^\mu (t, x)$\,, $\mu = 0\,, 1\,, \cdots, d-1$ that map the two-dimensional worldsheet $\Sigma$ to a $d$-dimensional target space $\CM$\,. We require that $X^\mu$ transform trivially under the worldsheet foliation-preserving diffeomorphisms, time-reversal symmetry, and Weyl transformation. Coupling $X^\mu$ to the worldsheet gravity introduced in \S\ref{sec:dwg} gives rise to the following sigma model action:
\be \label{eq:SX}
	S_X = \frac{1}{4\pi\alpha'} \int_\Sigma dt \, dx \, N \sqrt{\gamma} \, \Big\{ d_n X^\mu \, d_n X^\nu \, G_{\mu\nu} (X) - \gamma^{-1} \, \p_x X^\mu \, \p_x X^\nu \, H_{\mu\nu} (X) \Bigr\}\,,
\ee
where $[X^\mu] = 0$ and $G_{\mu\nu} (X)$ and $H_{\mu\nu} (X)$ are arbitrary functionals of $X^\mu$. All the terms in \eqref{eq:SX} are marginal. The background fields $G_{\mu\nu}$ and $H_{\mu\nu}$ are \emph{a priori} unrelated due to the lack of any boost symmetries that transform the worldsheet coordinates $t$ and $x$ into each other. The action \eqref{eq:SX} is invariant under the target space reparametrizations:
\be \label{eq:GHtransf}
	G'_{\mu\nu} (X') = \frac{\p X^\rho}{\p {X'}^\mu} \, \frac{\p X^\sigma}{\p {X'}^\nu} \, G_{\rho\sigma} (X)\,,
		\qquad
	H'_{\mu\nu} (X') = \frac{\p X^\rho}{\p {X'}^\mu} \, \frac{\p X^\sigma}{\p {X'}^\nu} \, H_{\rho\sigma} (X)\,,
\ee
i.e. both $G_{\mu\nu}$ and $H_{\mu\nu}$ transform as a symmetric two-tensor field. Under the condition that these two-tensor fields are non-degenerate, they are both qualified as metric fields. Therefore, the sigma model defined in \eqref{eq:SX} that consists of Lifshitz scalars at a $z=1$ Lifshitz point describes strings moving in a target space whose geometry is encoded in two metric fields. We will refer to this type of geometries as a \emph{bimetric geometry}. Intriguingly, even though the worldsheet lacks any boost symmetry, the target space geometry has the full Lorentzian boost symmetry. As we will see in later discussions, the bimetric theory that arises in our sigma model deviates drastically from the conventional bimetric formalism of massive gravity \cite{Hassan:2011zd} (see \S\ref{sec:lg}). Also note that the sigma model we consider here does not accommodate any Kalb-Ramond field, which breaks the time-reversal symmetry \eqref{eq:trsx}. In a variant of our sigma model that we will soon introduce in \S\ref{sec:trb}, the time-reversal symmetry is explicitly broken, and the same Kalb-Ramond term considered in relativistic string theory can be included. 

Although it is natural to write down the sigma model \eqref{eq:SX} that describes classical strings in a curved bimetric background, it would be instructive to have a more stringy explanation for how the curved spacetime arises.
In string theory, a curved spacetime emerges as a coherent state of strings, essentially by exponentiating the closed string vertex operators associated with the graviton and other massless states \cite{Green:1987sp, Polchinski:1998rq}. Similarly, in \eqref{eq:SX}, one may also consider a bimetric spacetime that perturbs around the flat metric, with
\be \label{eq:ghf}
	G_{\mu\nu} = \eta_{\mu\nu} + \tfrac{1}{2} \bigl( F_{\mu\nu} + f_{\mu\nu} \bigr)\,,
		\qquad%
	H_{\mu\nu} = \eta_{\mu\nu} + \tfrac{1}{2} \bigl( F_{\mu\nu} - f_{\mu\nu} \bigr)\,,
\ee
where both $F_{\mu\nu}$ and $f_{\mu\nu}$ are small. Focusing on states with a fixed spacetime  momentum $k_\mu$\,,  we require that
\be
	F_{\mu\nu} \propto e^{i k \cdot X} \, S_{\mu\nu}\,,
		\qquad%
	f_{\mu\nu} \propto e^{i k \cdot X} \, s_{\mu\nu}
\ee
are plane waves.
Then, the action \eqref{eq:SX} can be obtained by inserting an exponentiation of the following composite operators in the path integral:
\begin{subequations} \label{eq:vfvf}
\begin{align}
	V_F & = \frac{g^{}_F}{2} \int_\Sigma dt \, dx \, N \sqrt{\gamma} \, \Bigl( d_n X^\mu \, d_n X^\nu - \gamma^{-1} \, \p_x X^\mu \, \p_x X^\nu \Bigr) \, e^{i k \cdot X} \, S_{\mu\nu}\,, \label{eq:VF} \\[2pt]
	V_f & = \frac{g^{}_f}{2} \, \int_\Sigma dt \, dx \, N \sqrt{\gamma} \, \Bigl( d_n X^\mu \, d_n X^\nu + \gamma^{-1} \, \p_x X^\mu \, \p_x X^\nu \Bigr) \, e^{i k \cdot X} \, s_{\mu\nu}\,. \label{eq:vf}	
\end{align}
\end{subequations}
Such operators must respect the local worldsheet Weyl invariance, requiring which at the quantum level gives rise to consistency conditions on $S_{\mu\nu}$ and $s_{\mu\nu}$\,. One way to obtain these conditions is by demanding that the beta-functions of the sigma model vanish, such that the worldsheet theory is scale invariant at the quantum level. This RG analysis can be done on a flat worldsheet, and we will explore it in \S\ref{sec:rbsm} and argue in \S\ref{sec:lg} that $F_{\mu\nu}$ gives rise to a massless spin-two excitation. Nevertheless, it would be appealing to have a more direct derivation of such consistency conditions by requiring that the vertex operators \eqref{eq:vfvf} be free of Weyl anomalies. For this purpose, we need to understand how to renormalize the vertex operators on a curved worldsheet, which is beyond the current scope.   
Moreover, since our worldsheet theory \eqref{eq:SX} is generally \emph{not} conformal, it is not necessarily true that every physical state corresponds to a vertex operator.

The Nambu-Goto formalism of \eqref{eq:SX} can be obtained readily by integrating out the worldsheet gravitational fields $N$, $\gamma$\,, and $\chi$ in the path integral. Since there is no propagating degrees of freedom in our worldsheet gravity, we can eliminate $N$, $\gamma$\,, and $\chi$ by taking their equations of motion to be on-shell. There are two independent equations of motion from varying these worldsheet gravitational fields,
\be \label{eq:NxNgammasol}
	\frac{\chi}{\gamma} = \frac{G_{tx}}{G_{xx}}\,,
		\qquad
	\frac{N^2}{\gamma} = - \frac{\det G_{\alpha\beta}}{G_{xx} \, H_{xx}}\,,
		\qquad
	G_{\alpha\beta} = 
		\begin{pmatrix}
			G_{tt} & G_{tx} \\
			G_{xt} & G_{xx}
		\end{pmatrix},
\ee
where $G_{\alpha\beta} \equiv \p_\alpha X^\mu \, \p_\beta X^\nu \, G_{\mu\nu}$\,. To write down \eqref{eq:NxNgammasol}, we already assumed that both $G_{xx}$ and $H_{xx}$ are nonzero. Plugging \eqref{eq:NxNgammasol} into \eqref{eq:SX}, we find the analogue of the Nambu-Goto formalism of bosonic string theory,~\,\footnote{If $G_{xx} = 0$ and $H_{xx} \neq 0$\,, and assuming that $\gamma\,, \chi$\,, and $N$ are finite but nonzero, then integrating out $\chi$ in \eqref{eq:SX} requires $G_{tx} = 0$\,, and \eqref{eq:sng} becomes $S_\text{NG} = (2\pi\alpha')^{-1} \int dt \, dx \sqrt{- G_{tt} \, H_{xx}}$\,. Similarly, if $G_{xx} \neq 0$ and $H_{xx} = 0$\,, we are left with a constraint $\det G_{\alpha\beta} = 0$\,. The Nambu-Goto formalism \eqref{eq:sng} is also nonsingular if $H_{xx} / G_{xx}$ is held to be finite in the limit $H_{xx}\,, G_{xx} \rightarrow 0$\,. For example, in the relativistic limit, we have $H_{xx} \rightarrow G_{xx}$\,; no matter whether $H_{xx}$ and $G_{xx}$ are zero or not, \eqref{eq:sng} always reduces to the relativistic Nambu-Goto action with $S_\text{NG} \rightarrow (2\pi\alpha')^{-1} \int dt \, dx \sqrt{- \det G_{\alpha\beta}}$\,.
}
\be \label{eq:sng}
	S_\text{NG} = \frac{1}{2\pi\alpha'} \int dt \, dx \sqrt{\frac{H_{xx}}{G_{xx}}} \sqrt{- \det G_{\alpha\beta}}\,,
\ee
where $H_{xx} = \p_x X^\mu \, \p_x X^\nu \, H_{\mu\nu}$\,. In the special case when $G_{\mu\nu} = H_{\mu\nu}$\,, \eqref{eq:sng} reduces to the Nambu-Goto action that describes relativistic string theory.

The full worldsheet theory that couples the Lifshitz scalar $X^\mu$ to the topological Ho\v{r}ava gravity \eqref{eq:Sg} is given by
\begin{align} \label{eq:total}
\begin{split}
	S & = \frac{1}{4\pi\alpha'} \int_\Sigma dt \, dx \, N \sqrt{\gamma} \, \Big\{ d_n X^\mu \, d_n X^\nu \, G_{\mu\nu} (X) - \gamma^{-1} \, \p_x X^\mu \, \p_x X^\nu \, H_{\mu\nu} (X) \Bigr\} \\[2pt]
	 & \quad\, + \frac{1}{2\pi} \int_\Sigma dt \, dx \, N \sqrt{\gamma} \, \Bigl\{ (K^2 + d_n K) \, \Phi_\text{T} (X) - \gamma^{-1} (a^2 - \CD_{x} \, a) \, \Phi_\text{L} (X) \Bigr\},
\end{split}
\end{align}
where $\Phi_\text{T} (X)$ and $\Phi_\text{L} (X)$ are dilaton fields associated with the temporal and spatial derivative terms, respectively. Both the dilaton fields transform as target space scalars. In the special case when $G_{\mu\nu} = H_{\mu\nu}$ and $\Phi \equiv \Phi_\text{T} = \Phi_\text{L}$\,, the worldsheet action \eqref{eq:total} becomes
\begin{align} 
	S_\text{rel.} & = - \frac{1}{4\pi\alpha'} \int_\Sigma d^2x \, \sqrt{-h} \, h^{\alpha\beta} \p_\alpha X^\mu \, \p_\beta X^\nu \, G_{\mu\nu} (X) 
	 + \frac{1}{4\pi} \int_\Sigma dt \, dx \, \sqrt{-h} \, R(h)\, \Phi(X)\,,
\end{align}
with $x^\alpha = (t, x)$ and $h_{\alpha\beta}$ defined in \eqref{eq:hab}. This action gains emergent local Lorentzian symmetry and describes relativistic string theory.

Using the space diffeomorphism $\delta x = \xi (t, x)$ in \eqref{eq:fpxinf}, we fix the shift function $\chi$ to zero by choosing an appropriate value of $\xi$ in \eqref{eq:fpdNx}. However, the time diffeomorphism $\delta t = \zeta (t)$ is \emph{not} sufficient to fix $N$ and $
\gamma$ up to a conformal factor as in relativistic string theory. 
Plugging $\chi = 0$ into \eqref{eq:total}, we obtain the gauge-fixed action,
\begin{align} \label{eq:action}
\begin{split}
	S & = \frac{1}{4\pi \alpha'} \int_\Sigma dt \, dx \left\{ \frac{\sqrt{\gamma}}{N} \, \p_t X^\mu \, \p_t X^\nu \, G_{\mu\nu} (X) - \frac{N}{\sqrt{\gamma}} \, \p_x X^\mu \, \p_x X^\nu \, H_{\mu\nu} (X) \right\} \\[2pt]
	& \quad\, + \frac{1}{2\pi} \int_\Sigma dt \, dx \left\{ \p_t \biggl( \frac{\p_t \sqrt{\gamma}}{N} \biggr)\, \Phi_\text{T} (X) - \p_x \biggl( \frac{ \p_x N }{\sqrt{\gamma}} \biggr) \Phi_\text{L} (X) \right\} \,.
\end{split}
\end{align}

The flat target space limit is given by $G_{\mu\nu} = \eta_{\mu\nu}$\,, $H_{\mu\nu} =  c^2 \, \eta_{\mu\nu}$\,, and $\Phi_\text{T,\,L} = \Phi^{(0)}_\text{T,\,L}$\,, where $c$ is the worldsheet speed of light and $\Phi^{(0)}_\text{T,\,L}$ are constants. In this limit, \eqref{eq:action} becomes
\begin{align} \label{eq:flat}
\begin{split}
	S_\text{flat} & = \frac{1}{4\pi \alpha'} \int_\Sigma dt \, dx \Bigr( e \, \p_t X^\mu \, \p_t X_\mu - c^2 \, e^{-1} \, \p_x X^\mu \, \p_x X_\mu \Bigr) \\[2pt]
		& \quad\, + \frac{1}{2\pi} \int_\Sigma dt \, dx \left\{ \p_t \biggl( \frac{\p_t \sqrt{\gamma}}{N} \biggr)\, \Phi^{(0)}_\text{T} - \p_x \biggl( \frac{ \p_x N }{\sqrt{\gamma}} \biggr) \Phi^{(0)}_\text{L} \right\} \,,
\end{split}
\end{align}
where $e = \sqrt{\gamma} \, / \, N$.
The flat-spacetime action is invariant under the global Poincar\'{e} transformations,
\be
	\delta X^\mu = \Theta^\mu + \Lambda^\mu{}_\nu \, X^\nu,
\ee
which are isometries in the target space. The equation of motion from varying $X^\mu$ in $S_\text{flat}$ is
\begin{align}
	\p_t \lr e \, \p_t X^\mu \rr - c^2 \, \p_x \lr e^{-1} \, \p_x X^\mu \rr = 0\,.
\end{align}
Varying the geometrical data $e$ in $S_\text{flat}$ gives rise to part of the stress energy tensor,
\be \label{eq:T}
	T_1 \equiv \frac{1}{\alpha'} \lr e \, \p_t X^\mu \, \p_t X^\mu + c^2 \, e^{-1} \, \p_x X^\mu \, \p_x X_\mu \rr = 0\,.
\ee
The other component of the stress energy tensor comes from varying $\chi$ in the action \eqref{eq:SX}, followed by setting $\chi$ to zero, which gives
\be \label{eq:tT}
	T_2 \equiv \frac{2 \, e}{\alpha'} \, \p_t X^\mu \, \p_x X_\mu = 0\,.
\ee
The curvature terms in \eqref{eq:action} are total derivatives and do not make contribution to  the stress energy tensor components \eqref{eq:T} and \eqref{eq:tT}.
The equations \eqref{eq:T} and \eqref{eq:tT} are analogues of the Virasoro constraints in bosonic string theory. However, unlike the relativistic worldsheet theory, we do \emph{not} have enough diffeomorphism to fix $e$ to one.  Instead, the Virasoro-type constraint \eqref{eq:T} also involves $e (t, x)$\,.

If we further take the worldsheet to be flat and rescale $(t, x)$ such that $c=1$\,, then \eqref{eq:flat} becomes in form the same as the free relativistic worldsheet action 
\begin{align} \label{eq:relflat}
	S_\text{flat} \rightarrow \frac{1}{4\pi \alpha'} \int_\Sigma dt \, dx \, \Bigl( \p_t X^\mu \, \p_t X_\mu - \p_x X^\mu \, \p_x X_\mu \Bigr)
\end{align}
that underlies bosonic string theory. The distinction here is that, instead of fixing the full relativistic diffeomorphisms, we fixed part of the foliation-preserving diffeomorphisms followed by the special choice $N = \sqrt{\gamma} = 1$ to obtain \eqref{eq:relflat}. This implies that we have a rather different BRST symmetry and ghost action, which requires further studies. Therefore, the resemblance between \eqref{eq:relflat} and the relativistic string action is only at the classical level.

\subsection{Time-reversal breaking and trimetricity} \label{sec:trb}

We mentioned earlier that the sigma model defined in \eqref{eq:SX} does not accommodate any Kalb-Ramond term, which explicitly breaks the time-reversal symmetry \eqref{eq:trsx} that forbids any terms containing both space and time derivatives. However, the time-reversal symmetry is not an essential ingredient in our formalism. Explicitly breaking the time-reversal symmetry not only enriches the worldsheet topologies, but also introduces a plethora of extra Lagrangian terms that include the Kalb-Ramond term. Each of these extra terms is an infinite sum of a series of marginal operators, invariant under the foliation-preserving diffeomorphisms and the Weyl invariance,
\be \label{eq:SmX}
	S^\text{m.}_X = - \frac{1}{4\pi\alpha'} \int_{\Sigma} dt \, dx \, N \, d_n X^\mu \, \p_x X^\nu \Bigl\{ Q_{\mu\nu} (X) + B_{\mu\nu} (X) \Bigr\}\,,
\ee
where $Q_{\mu\nu}$ is a symmetric two-tensor, and $B_{\mu\nu}$ is an antisymmetric two-tensor that plays the role of a Kalb-Ramond field. The novel $Q$-term arises due to the lack of local boost symmetry on the worldsheet.

Without the time-reversal symmetry, there are three extra gravitational terms that contain both temporal and spatial derivatives and transform as scalars under the foliation-preserving diffeomorphisms \eqref{eq:fpdx},
\be \label{eq:L123}
	\frac{1}{\sqrt{\gamma}} \, \p_x K\,,
		\qquad
	\frac{1}{\sqrt{\gamma}} \, K \, a\,,
		\qquad
	\frac{1}{\sqrt{\gamma}} \, d_n a\,,
\ee
Here, $d_n$ acts on $a = a_x$ as
\be
	d_n a = n (a) - \frac{\p_x (\gamma^{-1} \, \chi)}{N} \, a\,,
\ee
where $n$ is the operator defined in \eqref{eq:nvf}. From \eqref{eq:L123}, we form two independent Weyl invariant combinations \cite{Arav:2014goa},
\begin{align} \label{eq:Sgm}
\begin{split}
	S^\text{mix}_\text{gr.} & = \frac{1}{{\tilde{\kappa}}^2} \int_\Sigma dt \, dx \, N \Bigl[ \lr \p_x K - K \, a \rr + \tilde{\alpha} \, d_n a \Bigr] \\[2pt]
		& = \frac{1}{{\tilde{\kappa}}^2} \int_\Sigma dt \, dx \, \Bigl[ \p_t \bigl( \tilde{\alpha} \, a \bigr) + \p_x \bigl( N K - \tilde{\alpha} \, \gamma^{-1} \chi \, a \bigr) \Bigr]\,,
\end{split}
\end{align}
whose integrand is a total derivative. At the fixed point $\tilde{\alpha} = 1$\,, $S^\text{mix}_\text{gr.}$ can be written in terms of the relativistic Zweibein field $e_\alpha{}^a$\,, $a = 0, 1$ \cite{Arav:2014goa},
\be \label{eq:Weylpartner}
	S^\text{mix}_\text{gr.} \rightarrow \frac{1}{2 \, {\tilde{\kappa}}^2} \int_\Sigma dt \, dx \, \det (e_\beta{}^c) \, \epsilon_{ab} \, \nabla^\alpha \omega_\alpha{}^{ab}\,,
\ee
where $\epsilon_{ab}$ is the two-dimensional Levi-Civita symbol and $\omega_\alpha{}^{ab}$ is the spin connection. The integrand of \eqref{eq:Weylpartner} takes the form of the Weyl partner of the Lorentz anomaly \cite{bertlmann2000anomalies}, which, for example, is generated in conformal field theories that contain $n_\text{L}$ holomorphic and $n_\text{R}$ anti-holomorphic fermions with a mismatch $n_\text{L} \neq n_\text{R}$ \cite{Chamseddine:1992ry}. 
Coupling the topological gravity \eqref{eq:Sgm} to the worldsheet scalar $X^\mu$, we uncover the following dilaton term:
\be \label{eq:SmPhi}
	S^\text{mix}_\Phi = \frac{1}{2\pi} \int_\Sigma dt \, dx \, N \, \Bigl\{ \bigl( \p_x K - K \, a \bigr) \, \Phi_1 (X) + d_n a \, \Phi_2 (X) \Bigr\}\,.
\ee

Finally, adding \eqref{eq:SmX} and \eqref{eq:SmPhi} to the time-reversal invariant action $S$ in \eqref{eq:total}, we find the extended sigma model that include all marginal terms,
\begin{align}
\begin{split}
	S_\text{extended} & = S + S^\text{mix}_X + S^\text{mix}_\Phi \\[2pt]
		& = \frac{1}{4\pi\alpha'} \int_\Sigma dt \, dx \, N \sqrt{\gamma} \, \Big\{ d_n X^\mu \, d_n X^\nu \, G_{\mu\nu} (X) - \gamma^{-1} \, \p_x X^\mu \, \p_x X^\nu \, H_{\mu\nu} (X) \Bigr\} \\[2pt]
		& \quad\, - \frac{1}{4\pi\alpha'} \int_{\Sigma} dt \, dx \, N \, d_n X^\mu \, \p_x X^\nu \Bigl\{ Q_{\mu\nu} (X) + B_{\mu\nu} (X) \Bigr\} \\[2pt]
	 	& \quad\, + \frac{1}{2\pi} \int_\Sigma dt \, dx \, N \sqrt{\gamma} \, \Bigl\{ (K^2 + d_n K) \, \Phi_\text{T} (X) - \gamma^{-1} \, (a^2 - \CD_{x} \, a) \, \Phi_\text{L} (X) \Bigr\} \\[2pt]
		& \quad\, + \frac{1}{2\pi} \int_\Sigma dt \, dx \, N \, \Bigl\{ \bigl( \p_x K  - K \, a \bigr) \, \Phi_1 (X) + d_n a \, \Phi_2 (X) \Bigr\}\,,
\end{split}
\end{align}
which describes sigma model in a curved background described by three symmetric two-tensor fields, $G_{\mu\nu}$\,, $H_{\mu\nu}$\,, and $Q_{\mu\nu}$\,, one antisymmetric two-tensor, $B_{\mu\nu}$\,, and four dilaton fields, $\Phi_\text{T}$\,, $\Phi_\text{L}$\,, $\Phi_1$\,, and $\Phi_2$\,. 
We will not consider the quantum mechanics of this more general sigma model that includes terms breaking the time-reversal symmetry. Instead, we will focus on the simpler time-reversal invariant case \eqref{eq:total} in the rest of this paper.

\section{Renormalization of Bimetric Sigma Models} \label{sec:rbsm}

We are now ready to compute the beta-functionals for the background fields in the sigma model \eqref{eq:action}, with the time-reversal symmetry imposed. We will not consider the beta-functionals of the dilaton fields in this paper, which require a more thorough understanding of the foliated worldsheet geometry and further examinations of the Weyl anomalies on a curved worldsheet. From now on, we simply take the flat worldsheet limit by setting $e = 1$\,, which does not affect the beta-functionals of the metric fields $G_{\mu\nu}$ and $H_{\mu\nu}$\,. We will therefore focus on the following renormalizable NLSM:
\begin{align} \label{eq:action2}
\begin{split}
	S_\text{E} & = \frac{1}{4\pi \alpha'} \int_\Sigma d\tau \, dx \, \Bigl\{ \p_\tau X^\mu \, \p_\tau X^\nu \, G_{\mu\nu} (X) + \p_x X^\mu \, \p_x X^\nu \, H_{\mu\nu} (X) \Bigr\}\,.
\end{split}
\end{align}
We have performed a Wick rotation $t = - i \tau$ in \eqref{eq:action} (together with $S = - i S_\text{E}$). Note that one is free to introduce a constant rescaling factor between the bimetric fields such that $G_{\mu\nu} \rightarrow G_{\mu\nu}$ and $H_{\mu\nu} \rightarrow c^2 \, H_{\mu\nu}$\,, by rescaling the worldsheet fields and coordinates.
Later in this section, we will also include the contributions from the dilaton fields to the beta-functionals of the metric fields. At the lowest order in $\alpha'$ that we are interested in, these contributions from dilatons are purely classical as in relativistic string theory. Since we are focusing on the RG flows, which only capture local properties of the system, we can drop any total derivative terms through the calculation. 

It is important to note that the action \eqref{eq:action2} is invariant under the following mapping:
\be \label{eq:swap}
	\tau \longleftrightarrow x\,,
		\qquad
	G_{\mu\nu} \longleftrightarrow H_{\mu\nu}\,.
\ee
This self-dual property has to be preserved in the resulting beta-functionals. We will use this duality frequently to guide our quantum calculations as well as a sanity check.

\subsection{Bimetric geometry in the target space}

As a preparation for quantum calculations, we investigate some essential ingredients of the target-space bimetric geometry here. Some of these ingredients have been introduced in \cite{Rosen:1940zza, Rosen:1940zz}. 

With respect to the metric fields $G_{\mu\nu}$ and $H_{\mu\nu}$\,, we introduce two Levi-Civita connections, $\nabla$ and $\Delta$\,, satisfying the compatibility conditions
$\nabla_{\!\rho} G_{\mu\nu} = 0$ and $\Delta_{\rho} H_{\mu\nu} = 0$\,, respectively.
Define the Christoffel coefficients $\Gamma^\rho{}_{\mu\nu}$ ($\Theta^\rho{}_{\mu\nu}$) associated with the connection $\nabla_{\!\mu}$ ($\Delta_\mu$),
\begin{subequations}
\begin{align}
	\Gamma^\rho{}_{\mu\nu} & = \frac{1}{2} \, G^{\rho\sigma} \bigl( \p_\mu G_{\nu\sigma} + \p_\nu G_{\mu\sigma} - \p_\sigma G_{\mu\nu} \bigr)\,, \\[2pt]
		\qquad%
	\Theta^\rho{}_{\mu\nu} & = \frac{1}{2} \, H^{\mu\nu} \bigl( \p_\mu H_{\nu\sigma} + \p_\nu H_{\mu\sigma} - \p_\sigma H_{\mu\nu} \bigr)\,. \label{eq:ThetaH}
\end{align}
\end{subequations}
Here, $G^{\mu\nu}$ ($H^{\mu\nu}$) is the inverse of $G_{\mu\nu}$ ($H_{\mu\nu}$).
The difference $S^\rho{}_{\mu\nu}$ between these two Christoffel symbols transforms as a $(1,2)$-tensor, with
\begin{align} \label{eq:defS}
\begin{split}
	S^\rho{}_{\mu\nu} = \Gamma^\rho{}_{\mu\nu} - \Theta^\rho{}_{\mu\nu} 
		& = \frac{1}{2} \, G^{\rho\sigma} \lr \Delta_\mu G_{\nu\sigma} + \Delta_\nu G_{\mu\sigma} - \Delta_\sigma G_{\mu\nu} \rr \\[2pt]
		& = - \frac{1}{2} \, H^{\rho\sigma} \lr \nabla_{\!\mu} H_{\nu\sigma} + \nabla_{\!\nu} H_{\mu\sigma} - \nabla_{\!\sigma} H_{\mu\nu} \rr.
\end{split}
\end{align}
We also define the Riemann curvature tensor $R^\rho{}_{\sigma\mu\nu}$ ($\Sigma^\rho{}_{\sigma\mu\nu}$) with respect to $\Gamma^\rho{}_{\mu\nu}$ ($\Theta^\rho{}_{\mu\nu}$), 
\begin{subequations}
\begin{align}
	R^\rho{}_{\sigma\mu\nu} & = \p_\mu \Gamma^\rho{}_{\nu\sigma} - \p_\nu \Gamma^\rho{}_{\mu\sigma} + \Gamma^\rho{}_{\mu\lambda} \, \Gamma^\lambda{}_{\nu\sigma} - \Gamma^\rho{}_{\nu\lambda} \, \Gamma^\lambda{}_{\mu\sigma}\,, \\[2pt]
	\Sigma^\rho{}_{\sigma\mu\nu} & = \p_\mu \Theta^\rho{}_{\nu\sigma} - \p_\nu \Theta^\rho{}_{\mu\sigma} + \Theta^\rho{}_{\mu\lambda} \, \Theta^\lambda{}_{\nu\sigma} - \Theta^\rho{}_{\nu\lambda} \, \Theta^\lambda{}_{\mu\sigma}\,.
\end{align}
\end{subequations}
The difference between the Riemann curvatures can be written in terms of the tensor $S^\rho{}_{\mu\nu}$ as
\begin{align}
\begin{split}
	R^\rho{}_{\sigma\mu\nu} - \Sigma^\rho{}_{\sigma\mu\nu} 
		& = \nabla_{\!\mu} S^\rho{}_{\nu\sigma} - \nabla_{\!\nu} S^\rho{}_{\mu\sigma} - S^\rho{}_{\mu\lambda} \, S^\lambda{}_{\nu\sigma} + S^\rho{}_{\nu\lambda} \, S^\lambda{}_{\mu\sigma} \notag \\[2pt]
		& = \Delta_{\mu} S^\rho{}_{\nu\sigma} - \Delta_{\nu} S^\rho{}_{\mu\sigma} + S^\rho{}_{\mu\lambda} \, S^\lambda{}_{\nu\sigma} - S^\rho{}_{\nu\lambda} \, S^\lambda{}_{\mu\sigma}\,.
\end{split}
\end{align}

With the above elements of bimetric geometry in hand, we are ready to use the covariant background field method to expand the worldsheet action \eqref{eq:action2}, as a preparation for evaluating the one-loop effective action on the worldsheet. We first choose $G_{\mu\nu}$ to be the reference metric, with respect to which the standard background field method can be applied. Consider a sufficiently small neighborhood $\CO$ of a point $X_0^\mu$ in the target space $\CM$\,. For an arbitrary point $X^\mu$ in $\CO$, and with respect to the reference metric field $G_{\mu\nu}$\,, there exists a unique geodesic interpolating between $X_0^\mu$ and $X^\mu$, parametrized by $Y^\mu (s)$\,, with an affine parameter $s \in [0, 1]$\,. The geodesic equation is
\be \label{eq:geod}
	\frac{d^2 Y^\mu (s)}{d s^2} + \Gamma^\mu{}_{\rho\sigma} \bigl(Y(s)\bigr) \, \frac{d Y^\rho (s)}{ds} \, \frac{d Y^\sigma (s)}{d s} = 0\,.
\ee  
We require that $Y^\mu$ satisfy the initial conditions $Y^\mu (0) = X_0^\mu$ and $Y^\mu (1) = X^\mu$. Define the tangent vector along the geodesic at $s=0$\,,
\be \label{eq:ellmu}
	p^\mu = \frac{dY^\mu(s)}{ds} \bigg|_{s=0}\,,
\ee
which constitutes a covariant quantum fluctuation. Note that $d/ ds$ commutes with the worldsheet derivative $\p_\alpha$\,, $\alpha = \tau, x$\,. It is also useful to introduce a covariant derivative $\nabla_{\!s}$ with respect to the affine parameter $s$ such that, for a covariant vector $u^\mu$ and a contravariant vector $v_\mu$\,,
\be \label{eq:nablas}
	\nabla_{\!s} u^\mu = \frac{du^\mu}{ds} + \Gamma^\mu{}_{\rho\sigma} \, u^\rho \, \frac{dY^\sigma}{ds} \,,
		\qquad
	\nabla_{\!s} v_\mu = \frac{dv_\mu}{ds} - \Gamma^\rho{}_{\mu\sigma} \, v_\rho \, \frac{dY^\sigma}{ds} \,.
\ee
Together with \eqref{eq:geod}, we have \cite{Howe:1986vm}
\begin{subequations} \label{eq:relations}
\begin{align}
	\nabla_{\!s} \, \p_\alpha Y^\mu & = \nabla_{\!\alpha} \frac{dY^\mu}{ds}\,,
		\qquad
	\nabla_{\!s} \frac{dY^\mu}{ds} = 0\,, \\[4pt]
	\bigl[ \nabla_{\!s} \,, \! \nabla_{\!\alpha} \bigr] \frac{dY^\mu}{ds} & = \p_\alpha Y^\nu R^\mu{}_{\rho\sigma\nu} (Y) \, \frac{dY^\rho}{ds} \, \frac{dY^\sigma}{ds}\,.
\end{align}
\end{subequations}
Furthermore, in parallel with \eqref{eq:nablas}, we define the covariant derivative $\Delta_s$ via
\be
	\Delta_s u^\mu = \frac{d v^\mu}{ds} + \Theta^\mu{}_{\rho\sigma} \, u^\rho \, \frac{dY^\sigma}{ds}\,,
		\qquad
	\Delta_s v_\mu = \frac{dv_\mu}{ds} - \Theta^\rho{}_{\mu\sigma} \, v_\rho \, \frac{dY^\sigma}{ds} \,.
\ee
The analogues of the relations in \eqref{eq:relations} are
\begin{subequations}
\begin{align}
	\Delta_s \, \p_\alpha Y^\mu & = \Delta_\alpha \frac{dY^\mu}{ds}\,,
		\qquad
	\Delta_s \frac{dY^\mu}{ds} = - S^\mu{}_{\rho\sigma} (Y) \, \frac{dY^\rho}{ds} \, \frac{dY^\sigma}{ds}\,, \\[4pt]
	\bigl[ \Delta_s\,, \Delta_\alpha \bigr] \frac{dY^\mu}{ds} & = \p_\alpha Y^\nu \, \Sigma^\mu{}_{\rho\sigma\nu} (Y) \, \frac{dY^\rho}{ds} \, \frac{dY^\sigma}{ds}\,.
\end{align}
\end{subequations}

Alternatively, we can choose $H_{\mu\nu}$ instead of $G_{\mu\nu}$ to be the reference metric, which must be equivalent when any physical observable is concerned. There also exists a unique geodesic interpolating between $X_0^\mu$ and $X^\mu$ and parametrized by ${Z}^\mu (r)$\,, defined with respect to $H_{\mu\nu}$ and an affine parameter $r \in [0,1]$\,, such that
\be \label{eq:geoZ}
	\frac{d^2 {Z}^\mu (r)}{d r^2} + \Theta^\mu{}_{\rho\sigma} \bigl(Z(r)\bigr) \, \frac{d {Z}^\rho (r)}{dr} \, \frac{d {Z}^\sigma (r)}{d r} = 0\,.
\ee  
The initial conditions are ${Z}^\mu (0) = X_0^\mu$ and ${Z}^\mu (1) = X^\mu$. The tangent vector along the geodesic at $r=0$ is
\be \label{eq:defk}
	q^\mu = \frac{d{Z}^\mu(r)}{dr} \bigg|_{r=0}\,.
\ee
Up to second order in $r$\,, the solution of $Z(r)$ to the geodesic equation \eqref{eq:geoZ} is
\be
	Z^\mu (r) = Z^\mu (0) + r \, q^\mu - \frac{1}{2} \, r^2 \, \Theta^\mu{}_{\rho\sigma} \, q^\rho \, q^\sigma + O(q^3)\,.
\ee
Setting $r = 1$\,, we find
\be \label{eq:Xk}
	X^\mu = X_0^\mu + q^\mu - \frac{1}{2} \, \Theta^\mu{}_{\rho\sigma} \, q^\rho \, q^\sigma + O(q^3)\,.
\ee
Similarly, from \eqref{eq:geod} we obtain
\be \label{eq:Xl}
	X^\mu = X_0^\mu + p^\mu - \frac{1}{2} \, \Gamma^\mu{}_{\rho\sigma} \, p^\rho \, p^\sigma + O(p^3)\,.
\ee
Using \eqref{eq:Xk} and \eqref{eq:Xl}, we find the following relation between the vectors $p^\mu$ and $q^\mu$ that are tangent at $X^\mu_0$ to the geodesics defined respectively with respect to the connections $\Gamma$ and $\Theta$\,:
\be \label{eq:qpSpp}
	q^\mu = p^\mu - \frac{1}{2} \, S^\mu{}_{\rho\sigma} \, p^\rho \, p^\sigma + O(p^3)\,.
\ee

In the following, we will use the definitions and relations above to perform a covariant expansion of \eqref{eq:action2} around $X_0^\mu$\,.

\subsection{Bimetric covariant expansions}

Now, we return to the action \eqref{eq:action2} and expand it around $X_0^\mu$, with respect to the reference metric $G_{\mu\nu}$ and the quantum fluctuation $p^\mu$ defined in \eqref{eq:ellmu}. To facilitate this calculation, we define the interpolating Lagrangian,
\be
	\CL (s) = \frac{1}{4\pi\alpha'} \Bigl\{ \p_\tau Y^\mu (s) \, \p_\tau Y^\nu (s) \, G_{\mu\nu} \bigl(Y(s)\bigr) + \p_x Y^\mu (s) \, \p_x Y^\nu (s) \, H_{\mu\nu} \bigl(Y(s)\bigr) \Bigr \}\,.
\ee
For conveniency, we set $\alpha' = 1 / (2 \pi)$\,. It follows that,
\begin{subequations}
\begin{align}
	\frac{d \CL(s)}{ds} \bigg|_{s=0} & = \frac{1}{2} \, \nabla_{\!s} \bigl( \p_\tau Y^\mu \, \p_\tau Y^\nu \, G_{\mu\nu} \bigr) + \frac{1}{2} \, \Delta_s \bigl( \p_x Y^\mu \, \p_x Y^\nu \, H_{\mu\nu} \bigr) \bigg|_{s=0} \notag \\[2pt]
		& = \p_\tau X_0^\mu \, \nabla_{\!\tau} p^\nu \, G_{\mu\nu} + \p_x X^\mu_0 \, \Delta_{x} p^\nu \, H_{\mu\nu}\,, \\[4pt]
	\frac{d^2 \! \CL (s)}{d s^2} \bigg|_{s=0} & = \nabla_{\!s} \biggl( \! \p_\tau Y^\mu \, \nabla_{\!\tau} \frac{dY^\nu}{ds} \, G_{\mu\nu} \! \biggr) + \Delta_s \biggl( \! \p_x Y^\mu \, \Delta_{x} \frac{dY^\nu}{ds} \, H_{\mu\nu} \! \biggr) \bigg|_{s=0} \notag \\[2pt]
		& = G_{\rho\sigma} \nabla_{\!\tau} p^\rho \, \nabla_{\!\tau} p^\sigma + \bigl( G_{\mu\lambda} \, R^\lambda{}_{\rho\sigma\nu} \, \p_\tau X^\mu_0  \, \p_\tau X_0^\nu \bigr) \, p^\rho \, p^\sigma \notag \\[2pt]
		& \quad + H_{\rho\sigma} \, \Delta_x p^\rho \, \Delta_x p^\sigma + \bigl( H_{\mu\lambda} \, \Sigma^\lambda{}_{\rho\sigma\nu} \, \p_x X^\mu_0  \, \p_x X_0^\nu \bigr) \, p^\rho \, p^\sigma \notag \\[2pt]
		& \quad - H_{\mu\lambda} \, \p_x X_0^\mu \, \Delta_x \bigl( S^\lambda{}_{\rho\sigma} \, p^\rho \, p^\sigma \bigr) \,. 
\end{align}
\end{subequations}
Therefore,
\be
	\CL (s) = \CL (0) + \frac{d \CL(s)}{ds} \bigg|_{s=0} s + \frac{1}{2} \frac{d^2 \! \CL (s)}{d s^2} \bigg|_{s=0} s^2 + O(s^3)\,.	
\ee
Setting $s=1$\,, we find that 
the covariant expansion of the action \eqref{eq:action2} with respect to the quantum fluctuation $p^\mu$ is given by
\begin{align} \label{eq:ceS}
	S_\text{E} = S^{(0)} + S^{(1)} + S^{(2)} + O(p^3)\,, 
\end{align}
where
\begin{subequations}
\begin{align}
	S^{(0)} & = \frac{1}{2} \int d\tau \, dx \, \bigl( G_{\mu\nu} \, \p_\tau X^\mu_0 \, \p_\tau X^\nu_0 + H_{\mu\nu} \, \p_x X^\mu_0 \, \p_x X^\nu_0 \bigr)\,, \\
	S^{(1)} & = - \int d\tau \, dx \, \bigl( G_{\mu\rho} \nabla_{\!\tau} \p_\tau X_0^\mu + H_{\mu\rho} \, \Delta_x \p_x X_0^\mu \bigr) \, p^\rho\,, \\[2pt]
	S^{(2)} & = \frac{1}{2} \int d\tau \, dx \, p^\rho \Bigl[ - \, G_{\rho\sigma} \nabla_{\!\tau}^2 + G_{\mu\lambda} \, R^\lambda{}_{\rho\sigma\nu} \, \p_\tau X_0^\mu \, \p_\tau X_0^\nu \notag \\[2pt]
		& \hspace{2.73cm} - H_{\rho\sigma} \, \Delta_x^2 + H_{\mu\lambda} \bigl( \Sigma^\lambda{}_{\rho\sigma\nu} \, \p_x X_0^\mu \, \p_x X_0^\nu + S^\lambda{}_{\rho\sigma} \, \Delta_x \p_x X_0^\mu \bigr) \Bigr] p^\sigma.
\end{align}
\end{subequations}
Note that we have performed a series of integrations by parts to get \eqref{eq:ceS}. The couplings are understood to be functionals of the background field $X_0$\,, satisfing the equation of motion
\be \label{eq:eom}
	\frac{\delta S^{(1)}}{\delta p^\rho} = 0 
		\quad \implies \quad
	G_{\rho\mu} \nabla_{\!\tau} \p_\tau X_0^\mu + H_{\rho\mu} \, \Delta_x \p_x X_0^\mu = 0\,.
\ee
Note that the covariant expansion \eqref{eq:ceS} defined with respect to the reference metric $G_{\mu\nu}$ can be transformed to be the one defined with respect to the reference metric $H_{\mu\nu}$ by applying the self-dual mapping \eqref{eq:swap}, supplemented with the following derived rules:
\be \label{eq:ellk}
	\nabla_{\!\mu} \longrightarrow \Delta_\mu\,,
		\qquad
	R^\mu{}_{\rho\sigma\nu} \longrightarrow \Sigma^\mu{}_{\rho\sigma\nu}\,,
		\qquad
	S^\mu{}_{\rho\sigma} \longrightarrow - S^\mu{}_{\rho\sigma}\,,
		\qquad 
	p^\mu \longrightarrow q^\mu\,.
\ee
Recall that $q^\mu$ is the quantum fluctuation defined in \eqref{eq:defk} with respect to the geodesic associated with the metric $H_{\mu\nu}$\,.   
This does not suffice our need for a manifestly self-dual formula: since the supplementary duality transformations in \eqref{eq:ellk} also act nontrivially on the quantum fluctuations $p^\mu$ and $q^\mu$, choosing to integrate out either $p^\mu$ or $q^\mu$ will lead to beta-functionals that do not manifest the self-dual transformations.
This drawback can be circumvented by taking the change of variables,
\be \label{eq:ell}
	p^\mu = \ell^\mu + \tfrac{1}{4} \, S^\mu{}_{\rho\sigma} \, \ell^\rho \, \ell^\sigma + O(\ell^3)\,.
\ee
Together with \eqref{eq:qpSpp}, we find
\begin{align}
	\ell^\mu & = p^\mu - \tfrac{1}{4} \, S^\mu{}_{\rho\sigma} \, p^\rho \, p^\sigma + O(p^3)
	= q^\mu + \tfrac{1}{4} \, S^\mu{}_{\rho\sigma} \, q^\rho \, q^\sigma + O(q^3)\,,
\end{align}
and it is evident that $\ell$ is invariant under the transformation \eqref{eq:ellk}. Integrating $\ell^\mu$ out will then lead to a self-dual one-loop effective action as desired. Comparing \eqref{eq:ell} with \eqref{eq:Xk} and the definition of $p^\mu$ in \eqref{eq:ellmu}, we find that $\ell^\mu$ is the tangent vector at $u = 0$ along the geodesic parametrized by an affine parameter $u$\,, satisfying the geodesic equation,
\be \label{eq:geou}
	\frac{d^2 U^\mu(u)}{d u^2} + \frac{1}{2} \, \Bigl[ \Gamma^\mu{}_{\rho\sigma} \bigl(U(u)\bigr) + \Theta^\mu{}_{\rho\sigma} \bigl(U(u)\bigr) \Bigr] \, \frac{dU^\rho(u)}{du} \, \frac{dU^\sigma(u)}{du} = 0\,,
\ee
with $U^\mu(0) = X_0^\mu$ and $U^\mu(1) = X^\mu$. The solution to the geodesic equation \eqref{eq:geou} is
\be \label{eq:Uusoln}
	U^\mu (u) = X_0^\mu + u \, \ell^\mu - \frac{1}{4} \, u^2 \, \bigl( \Gamma^\mu{}_{\rho\sigma} + \Theta^\mu{}_{\rho\sigma} \bigr) \, \ell^\rho \, \ell^\sigma + O(\ell^3)\,.
\ee	
Setting $u = 1$ in \eqref{eq:Uusoln}, we find
\be
	X^\mu = X_0^\mu + \ell^\mu - \frac{1}{4} \, \bigl( \Gamma^\mu{}_{\rho\sigma} + \Theta^\mu{}_{\rho\sigma} \bigr) \, \ell^\rho \, \ell^\sigma + O(\ell^3)\,.
\ee	
Comparing with \eqref{eq:Xl}, we recover the defining relation \eqref{eq:ell}. 

In terms of the new variable $\ell^\mu$\,, we find that \eqref{eq:ceS} becomes
\be \label{eq:ceaction}
	S_\text{E} = S^{(0)} + \tilde{S}^{(1)} + \tilde{S}^{(2)} + O(\ell^3)\,, 
\ee
with
\begin{subequations}
\begin{align}
	\tilde{S}^{(1)} & = - \int d\tau \, dx \, \bigl( G_{\mu\rho} \nabla_{\!\tau} \p_\tau X_0^\mu + H_{\mu\rho} \, \Delta_x \p_x X_0^\mu \bigr) \, \ell^\rho\,, \\[2pt]
	\tilde{S}^{(2)} & = \frac{1}{2} \int d\tau \, dx \, \ell^\rho \, \CO_{\rho\sigma} \, \ell^\sigma\!.
\end{align}
\end{subequations}
We have defined
\begin{align} \label{eq:defO}
\begin{split}
	\CO_{\rho\sigma} & \equiv - \, G_{\rho\sigma} \nabla_{\!\tau}^2 \hspace{0.5mm} - H_{\rho\sigma} \, \Delta_x^2 + V_{\rho\sigma}\,,
\end{split}
\end{align}
and
\begin{align} \label{eq:defV}
\begin{split}
	V_{\rho\sigma} & \equiv G_{\mu\lambda} \lr R^\lambda{}_{(\rho\sigma)\nu} \, \p_\tau X_0^\mu \, \p_\tau X_0^\nu - \tfrac{1}{2} \, S^\lambda{}_{\rho\sigma} \nabla_{\!\tau} \p_\tau X_0^\mu \rr \\[4pt]
	& \, + H_{\mu\lambda} \lr \Sigma^\lambda{}_{(\rho\sigma)\nu} \, \p_x X_0^\mu \, \p_x X_0^\nu + \tfrac{1}{2} \, S^\lambda{}_{\rho\sigma} \, \Delta_x \p_x X_0^\mu \rr.
\end{split}
\end{align}
Note that the operator $\CO_{\mu\nu}$ depends on the coordinates $(\tau, x)$ and the covariant derivatives $\nabla_{\!\tau}$ and $\Delta_x$\,. Moreover, $\CO_{\mu\nu} = \CO_{\nu\mu}$ and $V_{\mu\nu} = V_{\nu\mu}$\,. In the following, we will derive the one-loop effective action by integrating out the fluctuation $\ell^\mu$ in the path integral.  

\subsection{Heat kernel method for bimetric sigma models}

We are now ready to use the heat kernel method to compute the one-loop effective action associated with the sigma model \eqref{eq:action2}. We start with defining the effective action. Then, we will review some essential ingredients in the standard heat kernel method \cite{seeley1967complex, Gilkey:1975iq, Gusynin:1989ky} and discuss in detail how it is applied to our sigma model in a bimetric spacetime. 

\subsubsection{Heat kernel representation of the effective action}

We start with a quick review of the heat kernel method and derive the general form of the effective action, following \cite{seeley1967complex, Gilkey:1975iq, Gusynin:1989ky, Vassilevich:2003xt, Grosvenor:2021zvq} but with adaptions to the bimetric sigma models.
First, we define the associated path integral for the covariantly expanded action \eqref{eq:ceaction}, with respect to a reference metric $G_{\mu\nu} (X_0)$\,,
\be
	\CZ = \int d \ell^\mu \sqrt{- G(X_0)} \,\, \exp \biggl(- S_\text{E} (X) + \int d\tau \, dx \, J_\mu \, \ell^\mu \biggr)\,.
\ee
Even though we chose to define the measure in the path integral with respect to the reference metric $G_{\mu\nu}(X_0)$ with $G \equiv \det (G_{\mu\nu})$\,, it does \emph{not} give $G_{\mu\nu}$ any privileges in physical results as long as we integrate out all configurations of the quantum fluctuation. Later on, we will see explicitly that this choice preserves the self-dual property \eqref{eq:swap} in the final beta-functionals.
Choosing the background value $X_0$ such that $J_\mu = \delta \tilde{S}^{(1)} / \delta \ell^\mu$\,, in the semi-classical limit, the path integral is approximated by
\begin{align}
	\CZ 
		\sim \exp \Bigl( - S^{(0)} - \hbar \, \Gamma_\text{1-loop} + O(\hbar^2) \Bigr)\,,
\end{align}
where the one-loop effective action is
\begin{align} \label{eq:Gamma1}
\begin{split}
	\Gamma_\text{1-loop} & = \frac{1}{2} \, \tr \log \bigl( \CO_{\mu\rho} \, G^{\rho\nu} / m^2_\text{IR} \bigr) \\[2pt]
		& = - \frac{1}{2} \frac{d}{ds} \bigg|_{s=0} \frac{m_\text{IR}^{2s}}{\Gamma(s)} \int d\tau \, dx \int_0^\infty do \, o^{s-1} \, \CK_\mu{}^\mu \bigl(\{\tau, x\}, \{\tau, x\} \big| o\bigr)\,.
\end{split}
\end{align}
Here, $G^{\mu\nu}$ is the inverse of $G_{\mu\nu}$ and $m^{}_\text{IR}$ is an infrared (IR) cutoff. Note that $\CO_{\mu\nu}$ and $G^{\mu\nu}$ depend on the background value $X_0$ instead of $X$\,. We have defined the ``off-diagonal" heat kernel
\be \label{eq:hk}
	\CK_{\mu}{}^{\nu} \bigl(\tau, x\,; \tau_0\,, x_0 \big| o\bigr) = \bigl\langle \tau, x \bigl| \, \exp \bigl( - o \, \CO_{\mu\rho} \, G^{\rho\nu} \bigr) \, \bigr| \tau_0\,, x_0 \bigr\rangle\,,
\ee
which is a solution to the heat kernel equation, 
\be
	\bigl( \delta_\mu^\sigma \, \p_o + \CO_{\mu\rho} \, G^{\rho\sigma} \bigr) \, \CK_{\sigma}{}^{\nu} \bigl(\tau, x\,; \tau_0\,, x_0 \big| o\bigr) = 0\,.
\ee

It is useful to introduce the resolvent, 
$( \CO \, G^{-1} - \lambda \, \mathbb{1} )^{-1}$, 
using which we further rewrite the heat kernel \eqref{eq:hk} as
\be \label{eq:odk}
	\CK_{\mu}{}^{\nu} \bigl(\tau, x\,; \tau_0\,, x_0 \big| o\bigr) = \int_\CC \frac{i \, d\lambda}{2\pi} \, e^{-o \, \lambda} \, \CG_{\mu}{}^{\nu} (\tau, x\,; \tau_0\,, x_0 | \lambda)\,,
\ee
where $\CC$ is a contour that bounds the spectrum of the operator $\CO$ in the complex plane and is traversed in the counter-clockwise direction, and
\begin{align} \label{eq:propsymbol}
\begin{split}
	\CG_{\mu}{}^{\nu} \bigl(\tau, x\,; \tau_0\,, x_0 \big| \lambda \bigr) & \equiv \bigl\langle \tau, x \big| \, \bigl( \CO_{\mu\rho} \, G^{\rho\nu} - \lambda \, \delta_{\mu}^{\nu} \, \bigr)^{-1} \big| \tau_0\,, x_0 \bigr\rangle \\[2pt]
	& = \int \frac{d\omega \, d\kappa}{(2\pi)^2} \, e^{i \omega ( \tau - \tau_0 ) + i \kappa ( x - x_0 )} \, \sigma_{\mu}{}^{\nu} \bigl(\tau, x\,; \{ \tau_0\,, x_0 \}\,, \{\omega, \kappa\} | \lambda\bigr)\,.
\end{split}
\end{align}
We have introduced $\sigma_{\mu}{}^{\nu}$ as the \emph{symbol} of the resolvent, which essentially represents the Fourier modes of the resolvent.\,\footnote{On a curved worldsheet, there exists a covariant generalization of the phase function $e^{i \, \omega ( \tau - \tau_0 ) + i \, \kappa ( x - x_0 )}$ adapted to the foliation, defined using the symbolic calculus of pseudodifferential operators \cite{Grosvenor:2021zvq, Gusynin:1989ky, widom1980complete}. We will not need this covariant generalization of the phase function in this paper, but it will play an important role when it comes to the beta-functionals of the dilaton fields.} By definition,
\begin{align} \label{eq:id}
\begin{split}
	\bigl[ \CO_{\mu\rho} \bigl( \tau, x\,; \nabla_{\!\tau}, \Delta_x \bigr) \, G^{\rho\sigma} (\tau, x) - \lambda \, \delta_\mu^\sigma \bigr] \, \CG_\sigma{}^\nu \bigl(\tau, x\,; \tau_0\,, x_0 \big| \lambda \bigr) 
	= \delta(\tau - \tau_0) \, \delta (x - x_0) \, \delta_\mu^\nu\,.
\end{split}
\end{align}
Note that the derivatives only act on the first index of $\CG_\mu{}^\nu$\,. Plugging \eqref{eq:id} back into \eqref{eq:propsymbol}, we find
\begin{align} \label{eq:sigmaeq}
\begin{split}
	& \quad \bigl[ \CO_{\mu\rho} \bigl( \tau, x\,; \nabla_{\!\tau} + i \, \omega, \Delta_x + i \, \kappa \bigr) - \lambda \, G_{\mu\rho} (\tau, x) \bigr] \, \sigma^{\rho\nu} \bigl(\tau, x\,; \{ \tau_0\,, x_0 \}\,, \{\omega, \kappa\} | \lambda\bigr) \\[2pt]
	& = I_\mu{}^\nu \bigl( \tau, x\,; \tau_0\,, x_0 \bigr)\,.
\end{split}
\end{align}
We have defined
$\sigma^{\mu\nu} \equiv G^{\mu\rho} \, \sigma_{\rho}{}^\nu$\,.
Also note that only the covariant derivatives in $\CO$ that act directly on $\sigma^{\mu\nu}$ are shifted by $i \omega$ or $i \kappa$ in \eqref{eq:sigmaeq}. 
Moreoever, the bi-function $I_\mu{}^\nu \bigl( \tau, x\,; \tau^{}_0\,, x^{}_0 \bigr)$ is required to satisfy the following conditions in the coincidence limit $\tau \rightarrow \tau^{}_0$ and $x \rightarrow x^{}_0$\,:
\begin{align} \label{eq:coinI}
	I_\mu{}^\nu \bigl( \tau^{}_0\,, x^{}_0\,; \tau^{}_0\,, x^{}_0 \bigr) = \delta_\mu^\nu\,,
\end{align}
and
\be
	\nabla_{\!\tau}^k \, \Delta_x^\ell \, I_\mu{}^\nu \bigl( \tau, x\,; \tau^{}_0\,, x^{}_0 \bigr) \big|_{\tau = \tau^{}_0\,, \, x = x^{}_0} = 0\,,
		\qquad
	k + \ell \geq 1\,.
\ee
Consequently, quantities such as $\Delta_x \nabla_{\!\tau} I_\mu{}^\nu$ typically have a nonzero coincidence limit. 

The traced heat kernel that only depends on $\tau^{}_0$ and $x^{}_0$ has the following asymptotic expansion around $o \rightarrow 0^+$ \cite{seeley1967complex}:
\be \label{eq:CKE}
	\CK_\mu{}^\mu \bigl(\tau_0\,, x_0\,; \tau_0\,, x_0 \big| o \bigr) = \sum_{m=0}^\infty E_m (\tau_0\,, x_0) \, o^{\frac{m}{2}-1}\,.
\ee
To compute the heat kernel coefficients $E_m$\,, we expand the symbol $\sigma^{\mu\nu}$ as
\be
	\sigma = \sum_{m=0}^\infty \sigma_m\,,
\ee
where, in the coincidence limit, $\sigma_m$ is a homogeneous function of $\lambda$\,, $\tau$, and $x$\,, with
\be \label{eq:smh}
	\sigma_m \bigl( \tau_0\,, x_0\,; \{\tau_0\,, x_0\}\,, \{ b \, \omega, b \, \kappa \} \, \big| \, b^2 \, \lambda \bigr) = b^{-m-2} \, \sigma_m \bigl( \tau_0\,, x_0\,; \{\tau_0\,, x_0\}\,, \{ \omega, \kappa \} \, \big| \, \lambda \bigr)\,.
\ee
This motivates us to consider the following formal rescalings in \eqref{eq:sigmaeq}:
\be
	\omega \rightarrow b \, \omega\,,
		\quad
	\kappa \rightarrow b \, \kappa\,,
		\quad
	\lambda \rightarrow b^2 \, \lambda\,,
		\quad
	\sigma_m \rightarrow b^{-m-2} \, \sigma_m\,.
\ee
It follows that
\begin{align} \label{eq:prerecursion}
\begin{split}
	& \sum_{m=0}^\infty \bigl[ \CO_{\mu\rho} \bigl( \tau, x\,; \nabla_{\!\tau} + i \, b \, \omega, \Delta_x + i \, b \, \kappa \bigr) - b^2 \, \lambda \, G_{\mu\rho} (\tau, x) \bigr] \\[2pt]
	& \hspace{3cm} \times b^{-m-2} \, \sigma_m^{\rho\nu} \bigl(\tau, x\,; \{ \tau_0\,, x_0 \}\,, \{\omega, \kappa\} | \lambda\bigr)
	= I_\mu{}^\nu \bigl( \tau, x\,; \tau_0\,, x_0 \bigr)\,,
\end{split}
\end{align}
from which the coefficients $\sigma_m$ can be determined recusively by matching terms of different orders in $b$\,.
Taking the coincidence limit $\tau \rightarrow \tau^{}_0$ and $x \rightarrow x^{}_0$ in \eqref{eq:odk}, together with the rescalings,
\be
	\lambda \rightarrow o^{-1} \, \lambda\,,
		\qquad
	\omega \rightarrow o^{-1/2} \, \omega\,,
		\qquad
	\kappa \rightarrow o^{-1/2} \, \kappa\,,
\ee
we find
\begin{align} \label{eq:CKr}
	& \quad \CK_\mu{}^\nu \bigl(\tau^{}_0\,, x^{}_0\,; \tau^{}_0\,, x^{}_0 \big| o \bigr) \notag \\[2pt]
	& = o^{-2} \! \int_\CC \frac{i \, d\lambda}{2\pi} \, e^{-\lambda} \int \frac{d\omega \, d\kappa}{(2\pi)^2} \, \sigma_\mu{}^\nu \bigl(\tau^{}_0\,, x^{}_0\,; \{ \tau^{}_0\,, x^{}_0 \}\,, \{o^{-1/2} \, \omega\,, o^{-1/2} \, \kappa\} \big| \, o^{-1} \lambda\bigr)\,.
\end{align}
Plugging \eqref{eq:CKE} and \eqref{eq:smh} into \eqref{eq:CKr}, and matching terms of different orders in $o$\,, we find
\be \label{eq:Emsigma2}
	E_m (\tau, x) = \int \frac{d\omega \, d\kappa}{(2\pi)^2} \int_\CC \frac{i \, d\lambda}{2\pi} \, e^{-\lambda} \, G_{\mu\nu} (\tau, x) \, \sigma_m^{\mu\nu} \bigl(\tau, x\,; \{ \tau, x \}\,, \{ \omega, \kappa\} \big| \lambda\bigr)\,.
\ee

On the other hand, plugging \eqref{eq:CKE} into the effective action \eqref{eq:Gamma1}, regularized as
\begin{align}
	\Gamma_\text{1-loop} = - \frac{1}{2} \frac{d}{ds} \bigg|_{s=0} \frac{m_\text{IR}^{2s}}{\Gamma(s)} \int d\tau \, dx \int_{1/\Lambda^2}^\infty do \, o^{s-1} \, \CK_\mu{}^\mu \bigl(\{\tau, x\}, \{\tau, x\} \big| o\bigr)\,,
\end{align}
we find
\be \label{eq:g1loopex}
	\Gamma_\text{1-loop} = - \frac{1}{2} \int d\tau \, d x \, \biggl\{ E_0 \, \Lambda^2 +  2 \, E_1 \, \Lambda + E_2 \log \! \lr \frac{\Lambda^2}{m_\text{IR}^2} \rr + \text{finite} \biggr\}\,.
\ee
Clearly, the heat kernel coefficient $E_2$ contributes the log divergence in the one-loop effective action \eqref{eq:g1loopex}. This log divergence is associated with the beta-functionals of various couplings in the sigma model. The power-law divergences can be set to zero by choosing appropriate counterterms. 

\subsubsection{Solving the recursion relations}

Now, we compute the coincidence limit of the symbol $\sigma^{\mu\nu}_2$\,, from which we will be able to read off the beta-functionals by using \eqref{eq:Emsigma2}.
Using the explicit expression for $\CO$ in \eqref{eq:defO}, we find
\begin{align} \label{eq:Oshifted}
\begin{split}
	& \quad \CO_{\mu\nu} \bigl( \tau, x\,; \nabla_{\!\tau} + i \, b \, \omega, \Delta_x + i \, b \, \kappa \bigr) - b^2 \lambda \, G_{\mu\nu} \\[2pt]
	& = - G_{\mu\nu} \lr \nabla_{\!\tau} + i \, b \, \omega \rr^2 - H_{\mu\nu} \lr \Delta_x + i \, b \, \kappa \rr^2 + V_{\mu\nu} - b^2 \lambda \, G_{\mu\nu} \\[2pt]
	& = b^2 \CA_{\mu\nu} - 2 \, i \, b \lr \omega \, G_{\mu\nu} \nabla_{\!\tau} + \kappa \, H_{\mu\nu} \, \Delta_x \rr - \bigl( G_{\mu\nu} \nabla_{\!\tau}^2 + H_{\mu\nu} \, \Delta_x^2 - V_{\mu\nu} \bigr)\,,
\end{split}
\end{align}
where we defined 
\be \label{eq:AGH}
	\CA_{\mu\nu} \equiv \bigl( \omega^2 - \lambda \bigr) \, G_{\mu\nu} + \kappa^2 H_{\mu\nu}\,.
\ee
Plugging \eqref{eq:Oshifted} into \eqref{eq:prerecursion}, we find the following recursion relations:
\begin{subequations} \label{eq:recursionre}
\begin{align}
	\CA_{\mu\rho} \,\sigma_0^{\rho\nu} & = I_\mu{}^\nu, \\[2pt]
	\CA_{\mu\rho} \, \sigma_1^{\rho\nu} - 2 \, i \lr \omega \, G_{\mu\rho} \nabla_{\!\tau} + \kappa \, H_{\mu\rho} \, \Delta_x \rr \sigma_0^{\rho\nu} & = 0\,, \\[2pt]
	\CA_{\mu\rho} \, \sigma_2^{\rho\nu} - 2 \, i \lr \omega \, G_{\mu\rho} \nabla_{\!\tau} + \kappa \, H_{\mu\rho} \, \Delta_x \rr \sigma_1^{\rho\nu} - \lr G_{\mu\rho} \nabla_{\!\tau}^2 + H_{\mu\rho} \, \Delta_x^2 - V_{\mu\rho} \rr \sigma_0^{\rho\nu} & = 0\,.
\end{align}
\end{subequations}
There is also an infinite hierarchy of recursion relations that involve $\sigma_m$\,, $m > 2$\,, which we will \emph{not} need in this paper. We denote the coincidence limit of an object $Q (\tau, x\,; \tau^{}_0\,, x^{}_0)$ by $[Q] \equiv Q (\tau^{}_0\,, x^{}_0\,; \tau^{}_0\,, x^{}_0)$\,.  
Using \eqref{eq:coinI}, we find from \eqref{eq:recursionre} that
\begin{subequations}
\begin{align}
	[\sigma_0^{\mu\nu}] & = \CD^{\mu\nu}, \label{eq:sigma0cl} \\[2pt]
	[\sigma_1^{\mu\nu}] & = 2 \, i \, \CD^{\mu\rho} \Bigl( \omega \, G_{\rho\sigma} \, [\nabla_{\!\tau} \sigma_0^{\sigma\nu}] + \kappa \, H_{\rho\sigma} \, [\Delta_x \sigma_0^{\sigma\nu}] \Bigr)\,, \\[2pt]
	[\sigma_2^{\mu\nu}] & = \CD^{\mu\rho} \Bigl[2 \, i \bigl( \omega \, G_{\rho\sigma} [\nabla_{\!\tau} \sigma_1^{\sigma\nu}] + \kappa \, H_{\mu\rho} \, [\Delta_x \sigma_1^{\sigma\nu}] \bigr) + G_{\rho\sigma} [\nabla_{\!\tau}^2 \sigma_0^{\sigma\nu}] + H_{\rho\sigma} [\Delta_x^2 \sigma_0^{\sigma\nu}] \Bigr] \notag \\[2pt]
		& \quad\, - \CD^{\mu\rho} \, \CD^{\nu\sigma} \, V_{\rho\sigma}\,.  \label{eq:clsigma2}
\end{align}
\end{subequations}
Here, $\CD^{\mu\nu}$ is the inverse of $\CA_{\mu\nu}$\,, satisfying $\CD^{\mu\rho} \CA_{\rho\nu} = \delta^\mu_\nu$\,.
Recall that we are interested in the heat kernel coefficient $E_2$ defined in \eqref{eq:Emsigma2}, which is determined by $\sigma_2^{\mu\nu}$ and gives rise to the log divergence in the one-loop effective action \eqref{eq:g1loopex}. Further note that $E_2$ involves integrations over the frequency $\omega$ and momentum $\kappa$\,, and only receives nonzero contributions from terms in $[\sigma_2^{\mu\nu}]$ that are even in both $\omega$ and $\kappa$\,. According to \eqref{eq:clsigma2}, this implies that we only need to keep terms in $[\nabla_{\!\tau} \sigma_1^{\mu\nu}]$ that are odd in $\omega$ but even in $\kappa$\,, and terms in $[\Delta_x \sigma_1^{\sigma\nu}]$ that are odd in $\kappa$ and even in $\omega$\,. This observation brings some simplification in the calculation, which we will come back to momentarily. 

From the recursion relations in \eqref{eq:recursionre}, we compute the coincidence limits for expressions that involve $\sigma_m^{\mu\nu}$\,. We first introduce the simplifying notation,
\begin{subequations} \label{eq:notation}
\begin{align}
	\Omega^2 \equiv \omega^2 - \lambda\,,
		\qquad
	\bigl( H_\tau \bigr)^\mu{}_\nu & \equiv \CD^{\mu\rho} \, \nabla_{\!\tau} H_{\rho\nu}\,,
		\qquad
	\bigl( H_{\tau\tau} \bigr)^\mu{}_\nu \equiv \CD^{\mu\rho} \,  \nabla_{\!\tau}^2 H_{\rho\nu}\,, \\[2pt]
	\bigl( G_x \bigr)^\mu{}_\nu & \equiv \CD^{\mu\rho} \Delta_x G_{\rho\nu} \,,
		\qquad
	\bigl( G_{xx} \bigr)^\mu{}_\nu \equiv \CD^{\mu\rho} \Delta_{x}^2 G_{\rho\nu} \,.	
\end{align}
\end{subequations}
We also denote
\be
	\tilde{G}^\mu{}_\nu \equiv \CD^{\mu\rho} \, G_{\rho\nu}\,, 
		\qquad
	\tilde{H}^\mu{}_\nu \equiv \CD^{\mu\rho} \, H_{\rho\nu}\,.
\ee
In terms of the above definitions, it follows from \eqref{eq:recursionre} that 
\begin{subequations} \label{eq:cls}
\begin{align}
	[\nabla_{\!\tau} \sigma_0] & = - \kappa^2 \, H_\tau \, \CD\,, \\[2pt]
	[\Delta_x \sigma_0] & = - \Omega^2 \, G_x \, \CD\,, \\[2pt]
	[\nabla_{\!\tau}^2 \sigma_0] & = \bigl( 2 \, \kappa^4 \, H_{\tau}^2 - \kappa^2 \, H_{\tau\tau}  \bigr) \, \CD\,, \\[2pt]
	[\Delta_x^2 \sigma_0] & = \bigl( 2 \, \Omega^4 \, G_x^2  - \Omega^2 \, G_{xx}  \bigr) \, \CD\,, \\[2pt]
	[\sigma_1] & = - 2 \, i \, \bigl( \omega \, \kappa^2 \, \tilde{G} \, H_\tau + \kappa \, \Omega^2 \, \tilde{H} \, G_x \bigr) \, \CD\,, \\[2pt]
	[\nabla_{\!\tau} \sigma_1] & \sim 2 \, i \, \omega \bigl( \kappa^4 \, H_\tau \, \tilde{G} \, H_\tau + 2 \, \kappa^4 \, \tilde{G} \, H_\tau^2 - \kappa^2 \, \tilde{G} \, H_{\tau\tau} \bigr) \, \CD\,, \\[2pt]
	[\Delta_x \sigma_1] & \sim 2 \, i \, \kappa \bigl( \Omega^4 \, G_x \, \tilde{H} \, G_x + 2 \, \Omega^4 \, \tilde{H} \, G_x^2 - \Omega^2 \, \tilde{H} \, G_{xx}  \bigr) \, \CD\,.
\end{align}
\end{subequations}
Note that $\sigma_m$ and $\CD$ here carry raised indices, i.e. $\sigma_m = (\sigma_m^{\mu\nu})$ and $\CD = (\CD^{\mu\nu})$\,. 
We have omitted terms in the expressions of $[\nabla_{\!\tau} \sigma_1]$ that are even in $\omega$ and odd in $\kappa$\,; we also have omitted the terms in the expressions of $[\Delta_x \sigma_1]$ that are even in $\kappa$ and odd in $\omega$\,. These omitted terms do not make any contribution to the heat kernel coefficient $E_2$\,. We did not record the expressions $[\Delta_x \! \nabla_{\!\tau} \sigma_0]$ and $[\nabla_{\!\tau} \Delta_x \sigma_0]$\,, since they only contribute terms that we omit in $[\nabla_{\!\tau} \sigma_1]$ and $[\Delta_x \sigma_1]$\,. Finally, plugging \eqref{eq:cls} into \eqref{eq:clsigma2}, we find
\begin{align} \label{eq:sigma2result}
\begin{split}
	[\sigma_2] & = 2 i \bigl( \omega \, \tilde{G} \, [\nabla_{\!\tau} \sigma_1] + \kappa \, \tilde{H} \, [\Delta_x \sigma_1] \bigr) + \tilde{G} \, [\nabla_{\!\tau}^2 \sigma_0] + \tilde{H} \, [\Delta_x^2 \sigma_0] - \CD V \CD \\[2pt]
		& \sim \Bigl[ - 4 \, \omega^2 \, \kappa^4 \, \bigl( \tilde{G} \, H_\tau \bigr)^2 - 8 \, \omega^2 \, \kappa^4 \, {\tilde{G}}^2 \, H_\tau^2 + 4 \, \omega^2 \, \kappa^2 \, \tilde{G}^2 \, H_{\tau\tau} \\[2pt]
		& \qquad\!\! - 4 \, \kappa^2 \, \Omega^4 \, \bigl( \tilde{H} \, G_x \bigr)^2 - 8 \, \kappa^2 \, \Omega^4 \, {\tilde{H}}^2 \, G_x^2 + 4 \, \kappa^2 \, \Omega^2 \, {\tilde{H}}^2 \, G_{xx} \\[2pt]
		& \qquad\!\! + 2 \, \kappa^4 \, \tilde{G} \, H_\tau^2 - \kappa^2 \, \tilde{G} \, H_{\tau\tau} + 2 \, \Omega^4 \, \tilde{H} \, G_x^2 - \Omega^2 \, \tilde{H} \, G_{xx} - \CD \, V \Bigr] \, \CD\,.
\end{split}
\end{align}
Plugging the result of $[\sigma^{\mu\nu}_2]$ back into \eqref{eq:Emsigma2} and performing the integrals over $\omega$ and $\kappa$\,, the heat kernel coefficient $E_2$ can be derived. The exact result after performing these integrals is difficult to compute; however, as we will show in the following, when the difference between $G_{\mu\nu}$ and $H_{\mu\nu}$ is sufficiently small, the heat kernel coefficient can be computed order by order perturbatively.

\subsubsection{Perturbative expansion of the heat kernel coefficient} \label{sec:pehkc}

We now perform the frequency-momentum integrals in the expression for the heat kernel coefficient $E_2$ in \eqref{eq:Emsigma2}. From now on, we focus on the case where the difference between $G_{\mu\nu}$ and $H_{\mu\nu}$ is controlled by a sufficiently small parameter $\epsilon$\,, such that
\be \label{eq:GminusH}
	G_{\mu\nu} - H_{\mu\nu} = \epsilon \, f_{\mu\nu}\,.
\ee
In terms of $f$, the quantity $\CA_{\mu\nu}$ defined in \eqref{eq:AGH} can be written as
\be
	\CA_{\mu\nu} = \bigl( \omega^2 + \kappa^2 - \lambda \bigr) \, G_{\mu\nu} - \epsilon \, \kappa^2 \, f_{\mu\nu}\,.
\ee
Its inverse $\CD^{\mu\nu}$ can be expressed as a Taylor expansion with respect to $\epsilon$\,, given by
\be \label{eq:Dexp}
	\CD^{\mu\nu} = \sum_{n=0}^\infty \frac{\epsilon^n \, \kappa^{2n}}{\bigl(\omega^2 + \kappa^2 - \lambda \bigr)^{n+1}} \, ( g^n )^{\mu}{}_{\rho} \, G^{\rho\nu}\,,
\ee
where we defined the matrix $g = (g^\mu{}_\nu)$ with
\be \label{eq:defg}
	g^\mu{}_\nu \equiv G^{\mu\rho} \, f_{\rho\nu}\,.
\ee

The terms in \eqref{eq:sigma2result} that are of the zeroth order in $\epsilon$ are
\be
	[\sigma^{\mu\nu}_2]^{(0)} = - \frac{G^{\mu\rho} \, V_{\rho\sigma} \, G^{\sigma\nu}}{\omega^2 + \kappa^2 - \lambda}\,.
\ee
Note that we chose \emph{not} to expand $V_{\mu\nu}$ defined in \eqref{eq:defV} with respect to $\epsilon$\,. The contribution from $[\sigma_2^{\mu\nu}]^{(0)}$ to $E_2$ can be computed by using \eqref{eq:Emsigma2}, yielding
\begin{align}
	E_2^{(0)} & = - \, G^{\mu\nu} \, V_{\mu\nu} \int \frac{d\omega \, d\kappa}{(2\pi)^2} \int_\CC \frac{i \, d\lambda}{2\pi} \, \frac{e^{-\lambda}}{\omega^2 + \kappa^2 - \lambda}		
		= - \frac{1}{4\pi} \, G^{\mu\nu} \, V_{\mu\nu}\,.
\end{align}
Similarly, we find, up to the fourth order in $f$\,, 
\begin{subequations} \label{eq:E214}
\begin{align}
	E_2^{(1)} & = - \frac{1}{8\pi} \, G^{\mu\nu} \, g^{\rho}{}_\mu \, V_{\rho\nu} + \frac{1}{48 \pi} \, G^{\mu\nu} \! \lr \nabla_{\!\tau}^2 f_{\mu\nu} - 3 \, \Delta_x^2 f_{\mu\nu} \rr, \\[4pt]
	E_2^{(2)} & = - \frac{3}{32 \pi} \, G^{\nu\sigma} \, g^{\mu}{}_{\rho} \, g^{\rho}{}_{\sigma} \, V_{\mu\nu} \notag \\[2pt]
	 	& \quad + \frac{1}{64\pi} \, G^{\mu\rho} \, G^{\nu\sigma} \Bigl[ 2 f_{\mu\nu} \bigl( \nabla_{\!\tau}^2 f_{\rho\sigma} + \Delta_x^2 f_{\rho\sigma} \bigr) + \! \nabla_{\!\tau} f_{\mu\nu} \nabla_{\!\tau} f_{\rho\sigma} + 5 \, \Delta_x f_{\mu\nu} \, \Delta_x f_{\rho\sigma} \Bigr], \\[4pt]
	E_2^{(3)} & = - \frac{5}{64 \pi} \, G^{\mu\lambda} \, g^\rho{}_\sigma \, g^\sigma{}_\lambda \, g^\nu{}_{\rho} V_{\mu\nu} 
		+ \frac{1}{128\pi} \, G^{\rho\sigma} g^{\mu}{}_{\rho} \, g^\nu{}_\sigma \lr 5 \, \nabla_{\!\tau}^2 f_{\mu\nu} + \Delta_x^2 f_{\mu\nu} \rr \notag \\[2pt]
		& \quad\, + \frac{5}{128 \pi} \, G^{\rho\sigma} \, G^{\lambda\nu} \, g^{\mu}{}_\lambda \bigl( \nabla_{\!\tau} f_{\mu\rho} \nabla_{\!\tau} f_{\nu\sigma} - \Delta_x f_{\mu\rho} \, \Delta_x f_{\nu\sigma} \bigr)\,, \\[4pt]
	E_2^{(4)} & = - \frac{35}{512 \pi} \, G^{\nu\kappa} \, g^{\mu}{}_{\nu} \, g^{\lambda}{}_{\mu} \, g^{\rho}{}_{\lambda} \, g^{\sigma}{}_{\kappa} \, V_{\rho\sigma} \notag \\[2pt]
		& \quad + \frac{1}{1536\pi} \Bigl[ G^{\lambda\nu} \, g^{\mu}{}_{\lambda} \, g^\rho{}_\mu \, g^\sigma{}_\nu \, \bigl( 70 \, \nabla_{\!\tau}^2 f_{\rho\sigma} + 6 \, \Delta_x^2 f_{\rho\sigma} \bigr) \notag \\[2pt]
		& \qquad\qquad\qquad\hspace{5mm} + G^{\nu\lambda} \, G^{\sigma\kappa} \, g^{\mu}{}_{\lambda} \, g^{\rho}{}_{\kappa} \, \bigl( 28 \, \nabla_{\!\tau} f_{\mu\rho} \nabla_{\!\tau} f_{\nu\sigma} -12 \, \Delta_x f_{\mu\rho} \, \Delta_x f_{\nu\sigma} \bigr) \notag \\[2pt]
		& \qquad\qquad\qquad\hspace{5mm}  + G^{\rho\sigma} \, G^{\lambda\kappa} \, g^\mu{}_\lambda \, g^{\nu}{}_{\kappa} \, \bigl( 77 \, \nabla_{\!\tau} f_{\mu\rho} \, \nabla_{\!\tau} f_{\nu\sigma} - 3 \, \Delta_x f_{\mu\rho} \, \Delta_x f_{\nu\sigma} \bigr) \Bigr]\,.
\end{align}
\end{subequations}
This calculation can be straightforwardly extended to any higher orders in $\epsilon$\,.
The heat kernel coefficient is then
\begin{align} \label{eq:E2e4}
	E_2 = E_2^{(0)} + \epsilon \, E_2^{(1)} + \epsilon^2 \, E_2^{(2)} + \epsilon^3 \, E_2^{(3)} + \epsilon^4 \, E_2^{(4)} + O(\epsilon^5)\,.
\end{align}
Plugging \eqref{eq:E2e4} into \eqref{eq:g1loopex}, and rewriting all the terms that contain two spatial derivatives in terms of the reference metric $H_{\mu\nu}$ instead of $G_{\mu\nu}$\,, 
we find the following log-divergent  contribution to the one-loop effective action (up to boundary terms):
\begin{align} \label{eq:Gammalog}
	\Gamma_\text{1-loop}^\text{log} = \frac{1}{4\pi} \log \! \lr \frac{M}{m^{}_\text{IR}} \rr \int_\Sigma d\tau \, d x \, \Bigl( \CP_{\mu\nu}^G \, \p_\tau X^\mu_0 \, \p_\tau X^\nu_0 + \CP_{\mu\nu}^H \, \p_x X^\mu_0 \, \p_x X^\nu_0 \Bigr)\,.
\end{align}
Here, we have included the counterterms to cancel the dependence on the regulator $\Lambda$ and introduced the renormalization scale $M$\,.
To transcribe the expressions for $\CP_{\mu\nu}^G$ and $\CP_{\mu\nu}^H$\,, we define in parallel with $g^\mu{}_\nu$ in \eqref{eq:defg},
\be
	h^\mu{}_\nu \equiv H^{\mu\rho} \, f_{\rho\nu}\,.
\ee
Recall that $H^{\mu\nu}$ is the inverse of $H_{\mu\nu}$\,. Then,
\begin{subequations} \label{eq:MGH}
\begin{align}
	\CP^G_{\mu\nu} & = U^{\rho\sigma} \, R^\lambda{}_{\rho\sigma(\mu} \, G_{\nu)\lambda}
		+ \frac{1}{2} \, G_{\lambda(\mu} \nabla_{\!\nu)} \! \lr U^{\rho\sigma} \, S^\lambda{}_{\rho\sigma} \rr \notag \\[2pt]
		& \quad\, + \frac{\epsilon^2}{16} \, \nabla_{\!\mu} \, g^\rho{}_\sigma \nabla_{\!\nu} \, g^\sigma{}_{\rho}
		+ \frac{5 \, \epsilon^3}{32} \, g^\rho{}_\sigma \, \nabla_{\!\mu} g^\sigma{}_{\lambda} \, \nabla_{\!\nu} g^\lambda{}_{\rho} \notag \\[2pt]
		& \quad\, + \frac{7 \, \epsilon^4}{128} \, \bigl( 3 \, g^{\rho}{}_\sigma  \, g^\sigma{}_{\lambda} \, \delta^\kappa_\theta + 2 \, g^\rho{}_\lambda \, g^\kappa{}_\theta \bigr) \, \nabla_{\!\mu} g^\lambda{}_{\kappa}  \, \nabla_{\!\nu} g^\theta{}_{\rho}
		+ O(\epsilon^5)\,, \label{eq:MG} \\[10pt]
	\CP^H_{\mu\nu} & = U^{\rho\sigma}  \, \Sigma^\lambda{}_{\rho\sigma(\mu} \, H_{\nu)\lambda}
		- \frac{1}{2} \, H_{\lambda(\mu} \, \Delta_{\nu)} \! \lr U^{\rho\sigma} \, S^\lambda{}_{\rho\sigma} \rr \notag \\[2pt]
		& \quad\, + \frac{\epsilon^2}{16} \, \Delta_{\mu} \, h^\rho{}_\sigma\, \Delta_{\nu} \, h^\sigma{}_{\rho}
		- \frac{5 \, \epsilon^3}{32} \, h^\rho{}_\sigma \, \Delta_{\mu} h^\sigma{}_{\lambda} \, \Delta_{\nu} h^\lambda{}_{\rho} \notag \\[2pt]
		& \quad\, + \frac{7 \, \epsilon^4}{128} \, \bigl( 3 \, h^{\rho}{}_\sigma  \, h^\sigma{}_{\lambda} \, \delta^\kappa_\theta + 2 \, h^\rho{}_\lambda \, h^\kappa{}_\theta \bigr) \, \Delta_{\mu} h^\lambda{}_{\kappa}  \, \Delta_{\nu} h^\theta{}_{\rho} 
		+ O(\epsilon^5)\,, \label{eq:MH}
\end{align}
\end{subequations}
which contribute the beta-functionals of $G_{\mu\nu}$ and $H_{\mu\nu}$\,. We already substituted $V_{\mu\nu}$ with its definition in \eqref{eq:defV}. Note that the covariant derivatives of the metric fields can be expressed in terms of $S^\rho{}_{\mu\nu}$\,, which is defined in \eqref{eq:defS} as the difference between the Christoffel symbols,
\be
	\nabla_{\!\mu} \, g^\rho{}_\sigma = H^{\rho\lambda} \, \bigl( G_{\kappa\lambda} \, S^\kappa{}_{\sigma\mu} + G_{\kappa\sigma} \, S^\kappa{}_{\lambda\mu} \bigr)\,,
		\qquad
	\Delta_{\mu} \, h^\rho{}_\sigma = - G^{\rho\lambda} \, \bigl( H_{\kappa\lambda} \, S^\kappa{}_{\sigma\mu} + H_{\kappa\sigma} \, S^\kappa{}_{\lambda\mu} \bigr)\,.
\ee
We have used the following relation to write $\CP_{\mu\nu}^H$ in \eqref{eq:MH} in terms of the reference metric $H_{\mu\nu}$ instead of $G_{\mu\nu}$\,: 
\begin{align}
\begin{split}
	G^{\mu\nu} & = \Bigl[ \bigl( \mathbb{1} + \epsilon \, h \bigr)^{-1} \, H^{-1} \Bigr]^{\!\mu\nu} \\[2pt]
		& = \Bigl[ \delta^\mu_\rho - \epsilon \, h^\mu{}_\rho + \epsilon^2 \, \bigl( h^2 \bigr){}^\mu{}_\rho - \epsilon^3 \, \bigl( h^3 \bigr){}^\mu{}_\rho + \epsilon^4 \, \bigl( h^4 \bigr){}^\mu{}_\rho \Bigr] \, H^{\rho\nu} + O(\epsilon^5)\,.
\end{split}
\end{align}
In contrast, $\CP_{\mu\nu}^G$ in \eqref{eq:MG} is still written in terms of the reference metric $G_{\mu\nu}$\,.
We defined
\begin{align} \label{eq:Uex}
\begin{split}
	U^{\mu\nu} & \equiv \Bigl[ \delta_\rho^\mu + \tfrac{1}{2} \, \epsilon \, g^\mu{}_\rho + \tfrac{3}{8} \, \epsilon^2 \bigl( g^2 \bigr){}^\mu{}_\rho + \tfrac{5}{16} \, \epsilon^3 \bigl( g^3 \bigr){}^\mu{}_\rho + \tfrac{35}{128} \, \epsilon^4 \bigl( g^4 \bigr){}^\mu{}_\rho + O(\epsilon^5) \Bigr] \, G^{\rho\nu} \\[2pt]
		& = \Bigl[ \delta_\rho^\mu - \tfrac{1}{2} \, \epsilon \, h^\mu{}_\rho + \tfrac{3}{8} \, \epsilon^2 \bigl( h^2 \bigr){}^\mu{}_\rho - \tfrac{5}{16} \, \epsilon^3 \bigl( h^3 \bigr){}^\mu{}_\rho + \tfrac{35}{128} \, \epsilon^4 \bigl( h^4 \bigr){}^\mu{}_\rho + O(\epsilon^5) \Bigr] \, H^{\rho\nu}\,,
\end{split}
\end{align}
where the summands coincide with the lowest-order terms in the Taylor expansion with respect to $\epsilon$ of the matrix $\bigl( \mathbb{1} - \epsilon \, g \bigr)^{-1/2} \, G^{-1}$\,.
We show that $U = \bigl( \mathbb{1} - \epsilon \, g \bigr)^{-1/2} \, G^{-1}$ is valid to all orders in $\epsilon$ by following the steps detailed below. First, note that the $U$-dependent terms in \eqref{eq:MGH} are all from the last term in \eqref{eq:sigma2result}, i.e.,
\be \label{eq:IV}
	I^{\mu\nu}_V \equiv - \CD^{\mu\rho} \, V_{\rho\sigma} \, \CD^{\sigma\nu} \subset [\sigma_2^{\mu\nu}]\,,
\ee 
which contributes the following term in $E_2$\,:
\be
	\CI_V = \int \frac{d\omega \, d\kappa}{(2\pi)^2} \int_\CC \frac{i \, d\lambda}{2\pi} \, e^{-\lambda} \, G_{\mu\nu} \, I^{\mu\nu}_V \subset E_2\,.
\ee
Using the expression for $\CD^{\mu\nu}$ in \eqref{eq:Dexp}, we find
\begin{align} \label{eq:CIV}
	\CI_V 
		= - \frac{1}{4 \pi} \, U^{\mu\nu} \, V_{\mu\nu}\,,
\end{align}
where
\be \label{eq:Udef}
	U^{\mu\nu} = \Bigl[ \bigl( \mathbb{1} - \epsilon \, g \bigr)^{-1/2} \, G^{-1} \Bigr]^{\mu\nu} = \bigl( \sqrt{H^{-1} \, G} \, G^{-1} \bigr)^{\mu\nu}.
\ee
Further note that
\be
	U = \sqrt{H^{-1} \, G} \, G^{-1} = \frac{1}{\sqrt{G^{-1} \, H}} \, G^{-1} = \sqrt{G^{-1} \, H} \, H^{-1}\,,
\ee
which demonstrates that $U$ remains unchanged if $G$ and $H$ are swapped. The remaining perturbative expansion in \eqref{eq:MGH} can be extended to all orders in $\epsilon$ as well, leading to the exact heat kernel coefficient. Since this exact expression does not take any illuminating form, we refer the interested readers to Appendix \ref{app:chkc} for details.

Further note that the effective action \eqref{eq:Gammalog} is self-dual under the mapping \eqref{eq:swap},\footnote{Under the duality transformation \eqref{eq:swap}, we have $f \longleftrightarrow -f$\,, which induces $g \longleftrightarrow - h$\,.} which is a symmetry of the sigma model \eqref{eq:action2} that we started with. This is highly nontrivial since all the intermediate steps that eventually lead to the effective action are taken with respect to a reference metric $G_{\mu\nu}$\,. These intermediate steps (e.g., \eqref{eq:E214}) do \emph{not} manifest the self-duality transformation \eqref{eq:swap}. The fact that the final effective action \eqref{eq:Gammalog} is self-dual therefore acts as a strong sanity check of our calculation.

\subsection{Dilaton contributions} 

Up to now, we have been focusing on the flat worldsheet, where the contributions from the dilaton terms to the effective action are invisible. In this subsection, we revisit the dilaton terms on a curved worldsheet and evaluate the contributions from the dilatons to the RG flows of $G_{\mu\nu}$ and $H_{\mu\nu}$\,. 

Recall that the sigma model on a curved worldsheet is given by \eqref{eq:total}. 
In the flat worldsheet limit, the dilaton term vanishes; however, the dilaton term in \eqref{eq:total} still makes nontrivial contributions to the trace anomaly,
\begin{align} \label{eq:TPhi0}
\begin{split}
	T_\Phi = \p_\tau^2 \Phi_\text{T} + \p_x^2 \Phi_\text{L}		
		& = \p_\tau X^\mu \, \p_\tau X^\nu \, \nabla_{\!\mu} \nabla_{\!\nu} \Phi_\text{T} + \p_x X^\mu \, \p_x X^\nu \, \Delta_\mu \Delta_{\nu} \Phi_\text{L} \\[2pt]
		& \quad + \nabla_{\!\tau} \p_\tau X^\mu \, \nabla_{\!\mu} \Phi_\text{T} + \Delta_{x} \p_x X^\mu \, \Delta_{\mu} \Phi_\text{L}\,.
\end{split}
\end{align}
On the other hand, the trace of the stress energy tensor is related to the beta-functionals of $\beta_{\mu\nu}^G$ and $\beta_{\mu\nu}^H$ as
\be \label{eq:trace}
	T = - \frac{1}{2\alpha'} \bigl( \p_\tau X^\mu \, \p_\tau X^\nu \, \beta_{\mu\nu}^G + \p_x X^\mu \, \p_x X^\nu \, \beta_{\mu\nu}^H \bigr)\,.
\ee
For $T_\Phi$ in \eqref{eq:TPhi0} to be absorbed into the beta-functionals in \eqref{eq:trace}, we need to remove in \eqref{eq:TPhi0} the last two terms that depend on $\nabla_{\!\tau} \p_\tau X^\mu$ and $\Delta_x \p_x X^\mu$\,. This can be achieved by first
using the equation of motion \eqref{eq:eom} satisfied by the background field $X_0^\mu$\,, 
\be \label{eq:GHeom}
	G_{\mu\rho} \nabla_{\!\tau} \p_\tau X_0^\mu + H_{\mu\rho} \, \Delta_x \p_x X_0^\mu = 0\,,
\ee
which allows us to rewrite \eqref{eq:TPhi0} as
\begin{align} \label{eq:TPhi}
	T_\Phi 
		& = \p_\tau X^\mu \, \p_\tau X^\nu \, \nabla_{\!\mu} \nabla_{\!\nu} \Phi_\text{T} + \p_x X^\mu \, \p_x X^\nu \, \Delta_\mu \Delta_{\nu} \Phi_\text{L} \notag \\[2pt]
		& \quad + \nabla_{\!\tau} \p_\tau X^\mu \, \bigl( \nabla_{\!\mu} \Phi_\text{T} - G_{\mu\rho} \, H^{\rho\sigma} \Delta_\sigma \Phi_\text{L} \bigr)\,.
\end{align}
Setting the extra $\nabla_{\!\tau} \p_\tau X^\mu$ term to zero requires that the background fields satisfy 
\be \label{eq:GHPhi}
	G^{\mu\nu} \, \nabla_{\!\nu} \Phi_\text{T} = H^{\mu\nu} \, \Delta_\nu \Phi_\text{L}\,.
\ee
This condition is fullfilled, for example, when the dilaton fields are constant.
As a result,
\be
	T_\Phi = \p_\tau X^\mu \, \p_\tau X^\nu \, \nabla_{\!\mu} \nabla_{\!\nu} \Phi_\text{T} + \p_x X^\mu \, \p_x X^\nu \, \Delta_\mu \Delta_{\nu} \Phi_\text{L}\,.
\ee
Comparing with \eqref{eq:trace}, we find that the classical trace anomaly in \eqref{eq:TPhi0} can be absorbed into the beta-functionals $\beta_{\mu\nu}^G$ and $\beta_{\mu\nu}^H$ by adding the following terms: 
\begin{align} \label{eq:Weylanomaly}
	\beta_{\mu\nu}^{G} & \supset 2 \, \alpha' \, \nabla_{\!\mu} \nabla_{\!\nu} \Phi_\text{T}\,, 
		\qquad%
	\beta_{\mu\nu}^{H} \supset 2 \, \alpha' \, \Delta_\mu \Delta_{\nu} \Phi_\text{L}\,.
\end{align}
We emphsize that the extra condition \eqref{eq:GHPhi} has to be imposed, otherwise we would have the extra term $\nabla_{\!\tau} \p_\tau X^\mu$ in \eqref{eq:TPhi} that cannot be absorbed into any local counterterms.

\subsection{Beta-functionals and coupled Ricci flows}

In \eqref{eq:Gammalog} and \eqref{eq:Weylanomaly}, we derived the Weyl anomalies in the effective action that are generated by quantum corrections, which lead to the following beta-functionals:
\begin{subequations} \label{eq:betaGH}
\begin{align}
	\beta_{\mu\nu}^G & = - \alpha' \, \CP_{\mu\nu}^G + 2 \, \alpha' \, \nabla_{\!\mu} \nabla_{\!\nu} \Phi_\text{T} + O\bigl({\alpha'}^2\bigr) \notag \\[4pt]
		& = - \alpha' \, \Bigl\{ U^{\rho\sigma} R^\lambda{}_{\rho\sigma(\mu} \, G_{\nu)\lambda}
		+ \frac{1}{2} \, G_{\lambda(\mu} \nabla_{\!\nu)} \! \lr U^{\rho\sigma} \, S^\lambda{}_{\rho\sigma} \rr 
		- 2 \, \nabla_{\!\mu} \nabla_{\!\nu} \Phi_\text{T} \notag \\[2pt]
		& \qquad \quad + \frac{\epsilon^2}{16} \, \nabla_{\!\mu} \, g^\rho{}_\sigma \nabla_{\!\nu} \, g^\sigma{}_{\rho}
		+ \frac{5 \, \epsilon^3}{32} \, g^\rho{}_\sigma \, \nabla_{\!\mu} \, g^\sigma{}_{\lambda} \, \nabla_{\!\nu} \, g^\lambda{}_{\rho} \notag \\[2pt]
		& \qquad \quad + \frac{7 \, \epsilon^4}{128} \, \bigl( 3 \, g^{\rho}{}_\sigma  \, g^\sigma{}_{\lambda} \, \delta^\kappa_\theta + g^\rho{}_\lambda \, g^\kappa{}_\theta \bigr) \, \nabla_{\!\mu} \, g^\lambda{}_{\kappa}  \, \nabla_{\!\nu} \, g^\theta{}_{\rho}
		+ O(\epsilon^5) \Bigr\} + O\bigl({\alpha'}^2 \bigr)\,, \\[10pt]
	\beta_{\mu\nu}^H & = - \alpha' \, \CP_{\mu\nu}^H + 2 \, \alpha' \Delta_{\mu} \Delta_{\nu} \Phi_\text{L} + O\bigl({\alpha'}^2\bigr) \notag \\[4pt]
		& = - \alpha' \, \Bigl\{  U^{\rho\sigma} \, \Sigma^\lambda{}_{\rho\sigma(\mu} \, H_{\nu)\lambda}
		- \frac{1}{2} \, H_{\lambda(\mu} \, \Delta_{\nu)} \! \lr U^{\rho\sigma} \, S^\lambda{}_{\rho\sigma} \rr - 2 \, \Delta_\mu \Delta_{\nu} \Phi_\text{L} \notag \\[2pt]
		& \qquad \quad + \frac{\epsilon^2}{16} \, \Delta_{\mu} \, h^\rho{}_\sigma\, \Delta_{\nu} \, h^\sigma{}_{\rho}
		- \frac{5 \, \epsilon^3}{32} \, h^\rho{}_\sigma \, \Delta_{\mu} h^\sigma{}_{\lambda} \, \Delta_{\nu} h^\lambda{}_{\rho} \notag \\[2pt]
		& \qquad \quad + \frac{7 \, \epsilon^4}{128} \, \bigl( 3 \, h^{\rho}{}_\sigma  \, h^\sigma{}_{\lambda} \, \delta^\kappa_\theta + h^\rho{}_\lambda \, h^\kappa{}_\theta \bigr) \, \Delta_{\mu} h^\lambda{}_{\kappa}  \, \Delta_{\nu} h^\theta{}_{\rho} 
		+ O(\epsilon^5)	\Bigr\} + O\bigl({\alpha'}^2 \bigr)\,.
\end{align}
\end{subequations}
These beta-functionals need to be supplemented with the condition $G^{\mu\nu} \, \nabla_{\!\nu} \Phi_\text{T} = H^{\mu\nu} \, \Delta_\nu \Phi_\text{L}$ in \eqref{eq:GHPhi}.
We have recovered the dependence on $\alpha'$ here, which we set to $1/(2\pi)$ through the calculation. As a summary, we also collect below the definitions of various quantities that appear in \eqref{eq:betaGH}:
\begin{subequations}
\begin{align}
	f_{\mu\nu} & = G_{\mu\nu} - H_{\mu\nu}\,, 
		\qquad%
	g^\mu{}_\nu = G^{\mu\rho} \, f_{\rho\nu}\,, 
		\qquad%
	h^\mu{}_\nu = H^{\mu\rho} \, f_{\rho\nu}\,, \\[8pt]
	U & = \sqrt{H^{-1} \, G} \, G^{-1} = \sqrt{G^{-1} \, H} \, H^{-1}\,, \\[4pt]
	S^\rho{}_{\mu\nu} 		
		& = \frac{1}{2} \, G^{\rho\sigma} \lr \Delta_\mu G_{\nu\sigma} + \Delta_\nu G_{\mu\sigma} - \Delta_\sigma G_{\mu\nu} \rr = \frac{1}{2} \, G^{\rho\sigma} \lr \Delta_\mu f_{\nu\sigma} + \Delta_\nu f_{\mu\sigma} - \Delta_\sigma f_{\mu\nu} \rr \\[2pt]
		& = - \frac{1}{2} \, H^{\rho\sigma} \lr \nabla_{\!\mu} H_{\nu\sigma} + \nabla_{\!\nu} H_{\mu\sigma} - \nabla_{\!\sigma} H_{\mu\nu} \rr = \frac{1}{2} \, H^{\rho\sigma} \lr \nabla_{\!\mu} f_{\nu\sigma} + \nabla_{\!\nu} f_{\mu\sigma} - \nabla_{\!\sigma} f_{\mu\nu} \rr.
\end{align}
\end{subequations}
Recall that $R^\rho{}_{\mu\nu\sigma}$ ($\Sigma^\rho{}_{\mu\nu\sigma}$) and $\nabla_{\!\mu}$ ($\Delta_\mu$) are respectively the Riemann tensor and  covariant derivative defined with respect to the metric $G_{\mu\nu}$ ($H_{\mu\nu}$)\,. Under the duality map $G_{\mu\nu} \rightarrow H_{\mu\nu}$\,,
the beta-functional $\beta^G_{\mu\nu}$ is transformed into $\beta^H_{\mu\nu}$\,. To map out the full RG flow structure of the sigma model, however, we will also need to compute the beta-functionals for the dilaton fields $\Phi_\text{T,\,L}$ at the lowest order in $\alpha'$\,, which are, for example, important for determining the critical dimension (if such notion still exists) of the target space. However, the analysis of the dilaton beta-functionals requires a more thorough understanding of the worldsheet geometry and higher-loop calculation \cite{Callan:1989nz}, for which other techniques are needed. For example, evaluating the Weyl anomalies on a curved worldsheet with a foliation structure requires the method developed in \cite{Grosvenor:2021zvq}, which we leave for future studies.

The set of beta-functionals in \eqref{eq:betaGH} gives rise to a notion of coupled Ricci flows. In the limit $H_{\mu\nu} \rightarrow G_{\mu\nu}$\,, the condition \eqref{eq:GHPhi} gives $\nabla_{\!\mu}\Phi_\text{T} = \nabla_{\!\mu} \Phi_\text{L} \equiv \nabla_{\!\mu} \Phi$\,. Moreover, the two equations in \eqref{eq:betaGH} reduce to the following beta-functional in relativistic string theory:
\be \label{eq:Ricciflow}
	\frac{dG_{\mu\nu}}{d\iota} = \alpha' \bigl( R_{\mu\nu} + 2 \, \nabla_{\!\mu} \nabla_{\!\nu} \Phi \bigr) + O\bigl( {\alpha'}^2 \bigr)\,.
\ee
where $\iota = \ln M$\,, with $M$ the renormalization scale. Here, $R_{\mu\nu} \equiv R^\rho{}_{\mu\rho\nu}$ is the Ricci tensor defined with respect to the unique metric $G_{\mu\nu} = H_{\mu\nu}$\,. Note that \eqref{eq:Ricciflow} is Perelman's Ricci flow equation for the metric field \cite{Perelman:2006un}. The equations in \eqref{eq:betaGH} can be viewed as a generalization of the Ricci flow equation that governs the evolution of two different geometries in a coupled way.

\subsection{Bimetric dynamics and linearized gravity} \label{sec:lg}

Requiring Weyl invariance at the quantum level sets the beta-functionals in \eqref{eq:betaGH} to zero. This gives rise to the equations of motion that dictate the dynamics of the target space geometry,
\begin{subequations} \label{eq:speom}
\begin{align}
	0 & = U^{\rho\sigma} \, R^\lambda{}_{\rho\sigma(\mu} \, G_{\nu)\lambda}
		+ \frac{1}{2} \, G_{\lambda(\mu} \nabla_{\!\nu)} \! \lr U^{\rho\sigma} \, S^\lambda{}_{\rho\sigma} \rr - 2 \, \nabla_{\!\mu} \nabla_{\!\nu} \Phi_\text{T} \notag \\[2pt]
	& \quad\, + \frac{\epsilon^2}{16} \, \nabla_{\!\mu} \, g^\rho{}_\sigma \nabla_{\!\nu} \, g^\sigma{}_{\rho}
		+ \frac{5 \, \epsilon^3}{32} \, g^\rho{}_\sigma \, \nabla_{\!\mu} \, g^\sigma{}_{\lambda} \, \nabla_{\!\nu} \, g^\lambda{}_{\rho} \notag \\[2pt]
	& \quad\, + \frac{7 \, \epsilon^4}{128} \, \bigl( 3 \, g^{\rho}{}_\sigma  \, g^\sigma{}_{\lambda} \, \delta^\kappa_\theta + g^\rho{}_\lambda \, g^\kappa{}_\theta \bigr) \, \nabla_{\!\mu} \, g^\lambda{}_{\kappa}  \, \nabla_{\!\nu} \, g^\theta{}_{\rho}
		+ \cdots\,, \\[10pt]
	0 & = U^{\rho\sigma} \, \Sigma^\lambda{}_{\rho\sigma(\mu} \, H_{\nu)\lambda}
		- \frac{1}{2} \, H_{\lambda(\mu} \, \Delta_{\nu)} \! \lr U^{\rho\sigma} \, S^\lambda{}_{\rho\sigma} \rr - 2 \, \Delta_{\mu} \Delta_{\nu} \Phi_\text{L} \notag \\[2pt]
		& \quad\, + \frac{\epsilon^2}{16} \, \Delta_{\mu} \, h^\rho{}_\sigma\, \Delta_{\nu} \, h^\sigma{}_{\rho}
		- \frac{5 \, \epsilon^3}{32} \, h^\rho{}_\sigma \, \Delta_{\mu} h^\sigma{}_{\lambda} \, \Delta_{\nu} h^\lambda{}_{\rho} \notag \\[2pt]
		& \quad\, + \frac{7 \, \epsilon^4}{128} \, \bigl( 3 \, h^{\rho}{}_\sigma  \, h^\sigma{}_{\lambda} \, \delta^\kappa_\theta + h^\rho{}_\lambda \, h^\kappa{}_\theta \bigr) \, \Delta_{\mu} h^\lambda{}_{\kappa}  \, \Delta_{\nu} h^\theta{}_{\rho} 
		+ \cdots\,,
\end{align}
\end{subequations}
together with the condition $G^{\mu\nu} \, \nabla_{\!\nu} \Phi_\text{T} = H^{\mu\nu} \, \Delta_\nu \Phi_\text{L}$ in \eqref{eq:GHPhi}. Here, ``$\cdots$" denotes higher-order terms in $\alpha'$ and $\epsilon$\,. Of course, we also need the equations of motion from setting the beta-functionals of the dilaton fields to zero to determine the full dynamics of the target space geometry, which we leave for future studies.

In the single metric limit $G_{\mu\nu} \! = \!H_{\mu\nu}$\,, the spacetime equations of motion \eqref{eq:speom} reduce to
\be
	R_{\mu\nu} + 2 \, \nabla_{\!\mu} \nabla_{\!\nu} \Phi = 0\,.
\ee
In the case of a constant $\Phi$\,, this gives the Ricci flat condition that describes Einstein's gravity in absence of matter fields. In this sense, the equations of motion in \eqref{eq:speom} extend Einstein's gravity to a bimetric gravity. 

To further understand the dynamics of the bimetric geometry, we set the background dilaton fields to be constant and focus on the linearized part of the spacetime equations of motion in \eqref{eq:speom}. It is instructive to first linearize the beta-functionals in \eqref{eq:betaGH} and then impose the Weyl invariance to get the linearized spacetime equations of motion. 
We expand the metrics $G_{\mu\nu}$ and $H_{\mu\nu}$ around the Minkowskian metric $\eta_{\mu\nu}$ as in \eqref{eq:ghf}, with
\be
	G_{\mu\nu} = \eta_{\mu\nu} + \tfrac{1}{2} \bigl( F_{\mu\nu} + f_{\mu\nu} \bigr)\,,
		\qquad
	H_{\mu\nu} = \eta_{\mu\nu} + \tfrac{1}{2} \bigl( F_{\mu\nu} - f_{\mu\nu} \bigr)\,.
\ee
The infinitesimal gauge transformations of $F_{\mu\nu}$ and $f_{\mu\nu}$ are induced by \eqref{eq:GHtransf}, which at the linearized order read
\be \label{eq:deltaFf}
	\delta F_{\mu\nu} = 2 \, \bigl( \p_\mu \Xi_\nu + \p_\nu \Xi_\mu \bigr) + \cdots\,,
		\qquad
	\delta f_{\mu\nu} = 0 + \cdots\,,
\ee
where ``$\cdots$" denotes nonlinear terms.
For simplicity, we require the dilatons to be constant. Linearizing the beta-functionals $\beta^G_{\mu\nu}$ and $\beta^H_{\mu\nu}$ in $F_{\mu\nu}$ and $f_{\mu\nu}$\,, we find from \eqref{eq:betaGH} that
\begin{subequations} \label{eq:dgdh}
\begin{align}
	\beta^G_{\mu\nu} & = - \frac{\alpha'}{2} \, \eta^{\rho\sigma} \bigl( L^F_{\mu\rho\sigma\nu} 
	+ \tfrac{1}{2} \, \epsilon \, \p_\rho \p_\sigma f_{\mu\nu} \bigr) + \cdots \,, \\[2pt]
	\beta^H_{\mu\nu} & = - \frac{\alpha'}{2} \, \eta^{\rho\sigma} \bigl( L^F_{\mu\rho\sigma\nu} - \tfrac{1}{2} \, \epsilon \, \p_\rho \p_\sigma f_{\mu\nu} \bigr) + \cdots \,,
\end{align}
\end{subequations}
where 
\begin{subequations}
\begin{align}
	L^F_{\mu\rho\sigma\nu} & = \tfrac{1}{2} \, \bigl( \p_\rho \p_\sigma F_{\mu\nu} - \p_\rho \p_\nu F_{\mu\sigma} + \p_\mu \p_\nu F_{\rho\sigma} - \p_\mu \p_\sigma F_{\rho\nu} \bigr)
\end{align}
\end{subequations}
coincides with the linearized part of the Riemann tensor associated with the metric field $G_{\mu\nu} + H_{\mu\nu}$\,.
It follows from \eqref{eq:dgdh} that
\begin{align}
	\beta^F_{\mu\nu} & = - \alpha' \, \eta^{\rho\sigma} L^F_{\mu\rho\sigma\nu} + \cdots\,, 
		\qquad%
	\beta^f_{\mu\nu} = - \frac{1}{2} \, \alpha' \, \eta^{\rho\sigma} \p_\rho \p_\sigma f_{\mu\nu} + \cdots\,.
\end{align}
Imposing that the theory is free of Weyl anomalies sets both beta-functionals to zero and gives the linearized spacetime equations of motion
\be \label{eq:leom}
	\eta^{\rho\sigma} L^F_{\mu\rho\sigma\nu} = 0\,,
		\qquad
	\eta^{\rho\sigma} \p_\rho \p_\sigma f_{\mu\nu} = 0\,.
\ee
These expressions are invariant under the linearized infinitesimal gauge transformation \eqref{eq:deltaFf}. Therefore, at the linearized order, from $G_{\mu\nu}$ and $H_{\mu\nu}$ we form a massless spin-two gauge field $F_{\mu\nu}$ and a matrix $f_{\mu\nu}$ that satisfies the Klein-Gordon equation. 
The first linearized equation in \eqref{eq:leom} implies that the vertex operator $V_F$ in \eqref{eq:VF} is associated with a spin-two excitation. It is of immediate interest to verify the existence of such a spin-two excitation in the string spectrum by analyzing appropriate closed string vertex operators in flat spacetime. This will be an essential step for us to interpret the bimetric spacetime as emerging from a coherent state of strings, similar to how Einstein's gravity arises in relativistic string theory. This analysis requires future studies of the worldsheet dynamics.

We emphasize that the bimetric gravity that arises from our Lifshitz-type sigma model is in nature different from the usual bimetric formalism of massive gravity \cite{Hassan:2011zd}, already at the linearized order: in fact, none of the modes in \eqref{eq:leom} is massive!\,\footnote{See e.g. section 5.4 of \cite{deRham:2014zqa} for the mass eigenstates in the bimetric formulation of massive gravity.} Nevertheless, the linearized equations of motion \eqref{eq:leom} imply that our bimetric gravity is not necessarily against the usual no-go theorem that multiplets of interacting massless spin-two fields do not exist \cite{Boulanger:2000rq}: even though the target-space geometry is described by two metric fields $G_{\mu\nu}$ and $H_{\mu\nu}$\,, there is only one massless spin-two excitation formed by these two metric fields. In the single metric limit $H_{\mu\nu} \rightarrow G_{\mu\nu}$\,, i.e. $f_{\mu\nu} \rightarrow 0$\,, the extra modes are set to zero and we are left with Einstein's gravity. Our analysis here is only for the free theory, and it requires further analysis to determine whether the interacting theory is free of ghosts, which we leave for future work.\,\footnote{A preliminary analysis of the Lagrangian formalism for the linearized equations of motion in \eqref{eq:leom} shows that there are kinetic terms with a wrong sign, which could be problematic when interactions are included. Nevertheless, it is possible that the related ghost-like modes are removed by nonlinear Hamiltonian constraints in the full interacting theory. This endeavour requires first constructing an action principle and then performing a Hamiltonian analysis. On the other hand, as long as the worldsheet quantum field theory that underlies this bimetric gravity is well-defined, the worldsheet perspective will keep providing us with a solid foundation for future studies of this exotic gravity theory. It is therefore important to investigate whether the worldsheet theory is unitary.}

\section{Outlooks: Membranes at Quantum Criticality} \label{sec:omqc}

Since we have given up the worldsheet boost symmetry, strings no longer have any privileged position, and a broader spectrum of theories that are beyond string theory present. So far, we have been focusing on two-dimensional sigma models that describe strings propagating in a bimetric spacetime. In the following, we discuss natural generalizations to Lifshitz-type NLSMs that describe membranes at quantum criticality, following \cite{Horava:2008ih, ssl, as}.

We consider worldvolume theories described by sigma models at a Lifshitz point, in which case it is possible to construct quantum theories of membranes that are potentially perturbatively defined. A quantum theory of membranes of this type is introduced in \cite{Horava:2008ih} and has been preluded in \S\ref{sec:intro}, where the worldvolume theory of the sigma model is described by a three-dimensional quantum field theory at a $z=2$ Lifshitz point, coupled to worldvolume Ho\v{r}ava gravity. Due to the foliation structure induced by the foliation-preserving diffeomorphism on the worldvolume, the summation over the three-manifold can be consistently restricted to be over a specific class of foliated manifolds, whose spatial leaves are Riemann surfaces. The spatial topology in the foliated manifold changes when the membranes interact with each other.  When the three-dimensional sigma model satisfies the so-call ``detailed balance condition," i.e., the time-evolution of the worldsheet gravitational and matter fields is governed by a gradient flow generated by some Euclidean action principle, an intriguing connection to string theory can be established: as shown in \cite{Horava:2008ih}, the quantum theory of membranes that satisfies the detailed balance condition, and with compact spatial topology $\Sigma_h$\,, has the property that its ground-state wavefunction reproduces the partition function of bosonic string theory on the worldsheet $\Sigma_h$\,. Moreover, the three-dimensional sigma model that describes membranes at quantum criticality inherits the RG properties of the two-dimensional relativistic sigma model that underlies bosonic string theory. As a result, the appropriate spacetime geometry coupled to the membranes is the same as in string theory, and the dynamics of the spacetime geometry is described by Einstein's gravity.
 
It is, however, intriguing to consider sigma models that generalize the theory of membranes at quantum criticality in \cite{Horava:2008ih} by relaxing the detailed balance condition. This will allow us to probe more exotic spacetime geometries. We will discuss different generalizations below. An especially interesting example has been introduced in \cite{ssl, as}, where a three-dimensional NLSMs with Aristotelian supersymmetry is formulated, presenting a nonrelativistic ultra-violet (UV) completion of three-dimensional relativistic $\CN=1$ supersymmetric sigma models. This Lifshitz-type NLSM provides a manageable candidate for a quantum membrane theory in a bimetric spacetime. 

\subsection{Bosonic membranes and \texorpdfstring{$O(N)$}{ON} nonlinear sigma model}

For simplicity, we will impose the worldvolume time-reversal symmetry throughout this section.
In the bosonic case, the generalization to theories of membranes leads to a proliferation of terms. To construct such membrane theories, we introduce the coordinates $(t, x^i)$\,, $i = 1, 2$ on the worldvolume. The engineering scaling dimensions at a $z=2$ Lifshitz point are
\be
	[t] = -1\,,
		\qquad
	[\mathbf{x}] = - \frac{1}{2}\,.
\ee
We parametrize the target space by $X^\mu$\,, $\mu = 0, 1, \cdots, d-1$\,, whose scaling dimension is zero. We classify all the marginal terms in the most general bosonic NLSM on a flat worldvolume:
\begin{align} \label{eq:Sbrane}
\begin{split}
	S_\text{brane} = \frac{T_2}{2} \int dt \, d^2 \mathbf{x} \, \Bigl\{ \p_t X^\mu \, \p_t X^\nu \, G_{\mu\nu} (X) & - \Delta^{\!i} \p_i X^\mu \, {\tilde{\Delta}}^{\!j} \p_j X^\nu \, Q_{\mu\nu}(X) \\[2pt]
	& - \p_i X^\mu \, \p_i X^\nu \, \p_j X^\rho \, \p_j X^\sigma \, T_{\mu\nu\rho\sigma}(X) \Bigr\}\,,
\end{split}
\end{align}
where the covariant derivative $\Delta_i$ ($\tilde{\Delta}_i$) is the pullback of $\Delta_\mu$ ($\tilde{\Delta}_\mu$) that is compatible with a metric field $H_{\mu\nu}(X)$ ($\tilde{H}_{\mu\nu} (X)$). Here, $G_{\mu\nu}$\,, $H_{\mu\nu}$\,, $\tilde{H}_{\mu\nu}$\,, and $Q_{\mu\nu}$ are symmetric and transform as two-tensors and $T_{\mu\nu\rho\sigma}$ transforms as a four-tensor with respect to spacetime diffeomorphisms. 
The action $S_\text{brane}$ is invariant under reparametrizations of the target space coordinates $X^\mu$. At the detailed balance, we have $G_{\mu\nu} = H_{\mu\nu} = \tilde{H}_{\mu\nu} = Q_{\mu\nu}$ and $T_{\mu\nu\rho\sigma} = 0$\,, and the beta-functional of $G_{\mu\nu}$ is the usual Ricci flow equation as in \eqref{eq:Ricciflow} (with the dilaton field set to zero). However, without the detailed balance condition, the spacetime geometry in \eqref{eq:Sbrane} is quite intricate and described by multiple tensorial fields. The sigma model \eqref{eq:Sbrane} can also be consistently coupled to the worldvolume Weyl-invariant Ho\v{r}ava gravity, which does not possess any propagating gravitational degree of freedom \cite{Horava:2008ih}. 

In the simple case when the target space is an $(d-1)$-sphere $S^{d-1}$\,, the sigma model in \eqref{eq:Sbrane} still deviates from its relativistic counter-partner in an interesting way. The marginal Lagrangian terms invariant under the target space $O(d)$ symmetry are \cite{Anagnostopoulos:2010gw, lnlsm, Yan:2017mse}\,\footnote{Also see \cite{Griffin:2013dfa} for studies of spontaneous symmetry breaking in Lifshitz-type NLSMs.}
\begin{align} \label{eq:Ssphere}
\begin{split}
	S_\text{sphere} & = \frac{T_2}{2} \int dt \, d^2 \mathbf{x} \, \Bigl\{ \p_t X^I \, \p_t X^J \, G_{IJ} - \zeta^2 \, \nabla^i \p_i X^I \, \nabla^j \p_j X^J \, G_{IJ} \\[2pt]
	& \qquad\quad\, -  \alpha_1 \, \bigl( \p_i X^I \, \p_i X^J \, G_{IJ} \bigr)^2 - \alpha_2 \, \bigl( \p_i X^I \, \p_j X^J \, G_{IJ} \bigr) \, \bigl( \p_i X^K \, \p_j X^L \, G_{KL} \bigr)  \Bigr\}\,,
\end{split}
\end{align}
where $I = 1, \, \cdots, \, d$ and $G_{IJ}$ is the round metric on $S^{N-1}$, 
\be
	G_{IJ} (X) = \delta_{IJ} + \frac{X^I \, X^J}{1 - X^K \, X^K}\,.
\ee
The RG flows of the NLSM in \eqref{eq:Ssphere} have been evaluated in \cite{lnlsm, Yan:2017mse}. Intriguingly, projected in the $\alpha_1$-$\alpha_2$ plane, the RG flows possess multiple fixed points, depending on the value of $d$\,. In particular, there is one RG fixed point that is independent of $d$ at $\alpha_1 = \alpha_2 = 0$\,. At this common fixed point in the $\alpha_1$-$\alpha_2$ plane, the theory exhibits  the detailed balance properties. The beta-function calculation of the $O(d)$ NLSM \eqref{eq:Ssphere} is already rather involved, due to the presence of the terms that are quartic in spatial derivatives. When it comes to the NLSM \eqref{eq:Sbrane} in general target spaces, the proliferation of tensorial structures in spacetime poses more challenges, which might require new techniques to tackle with.

\subsection{Supermembranes in a bimetric spacetime}

Just like the case of the $O(N)$ NLSM, imposing the detailed balance condition on $S_\text{brane}$ in \eqref{eq:Sbrane} significantly simplifies the quartic derivative terms. At the meantime, the detailed balance condition also forces $G_{\mu\nu} = H_{\mu\nu}$\,. Nevertheless, it is indeed possible to simplify the quartic derivative terms but still allow a bimetric structure in the target space. This is achieved in \cite{ssl, as} by introducing Aristotelian supersymmetry in three dimensions, generated by two real supercharges $Q_1$ and $Q_2$ that form a multiplet $Q_\alpha$\,, $\alpha = 1, 2$\,. 

We follow closely \cite{as} below and review the basic ingredients in the construction of the theory of supermembranes in a bimetric spacetime. The same superspace formalism for Lifshitz-type $O(N)$ sigma models can be found in \cite{Gomes:2016tus}. We start with some conventions. Define the two-dimensional Dirac gamma matrices,
\be
	\rho = 
		\begin{pmatrix}
			0 & \,\, 1 \\
			-1 & \,\, 0
		\end{pmatrix}\,,
	\qquad
	\rho^{}_1 = 
		\begin{pmatrix}
			0 & \,\,\,\, 1 \\
			1 & \,\,\,\, 0
		\end{pmatrix}\,,
	\qquad
	\rho^{}_2 = 
		\begin{pmatrix}
			1 & \,\, 0 \\
			0 & \,\, -1
		\end{pmatrix}.
\ee
For any given Grassmannian variable $\chi_\alpha$\,, we define its conjugate $\overline{\chi} = i \, \chi^\intercal \rho$\,.
Consider the superalgebra in which $Q_\alpha$ satisfies the anti-commutative relation
\be \label{eq:susyalg}
	\{ \overline{Q}^{}_\alpha\,, Q^{}_\beta \} = 2 \, i \, \rho^{}_{\alpha\beta} \, P_0\,, 
\ee
where $P_0$ is the energy generator. The relation in \eqref{eq:susyalg} can be deformed by a relevant term such that 
\be \label{eq:QQc}
	\{ \overline{Q}\,, Q \} = 2 \, i \bigl( \rho \, P_0 + c \, \rho^{}_i P_i \bigr)\,,
\ee
where $c$ is a dimensionful coupling (that plays the role of speed of light) from the perspective of the UV Aristotelian observer and $P_i$ is the spatial momentum generator. Together with temporal and spatial translations and spatial rotational transformations (but without any boost transformation), we form the algebra of Aristotelian supersymmetry \cite{ssl, as} (also see \cite{Frenkel:2020djn} for related discussions). Since we are zooming in around the UV $z=2$ Lifshitz point here, we will tune the coupling $c$ to zero in the following discussion.

Next, we introduce worldvolume fermions $\Psi^\mu_\alpha = (\Psi^\mu_1\,, \Psi^\mu_2)^\intercal$ together with an auxiliary field $B^{\mu}$\,. The fermionic symmetry transformations generated by $Q_\alpha$ and parametrized by the Grassmannian number $\epsilon = (\epsilon_1\,, \epsilon_2)^\intercal$ are
\begin{subequations} \label{eq:susytrsf}
\begin{align}
	\delta_\epsilon X^\mu & = \overline{\epsilon} \, \Psi^\mu\,, \\[2pt]
	\delta_\epsilon \Psi^\mu & = \bigl( - \p_t X^\mu \rho + B^{\mu} \bigr) \epsilon, \\[2pt]
	\delta_\epsilon B^{\mu} & = \overline{\epsilon} \, \rho \, \p_t \Psi^\mu\,.
\end{align}
\end{subequations}
In the superspace formalism, in addition to the coordinates $(t\,, x^i)$\,, we introduce two real Grassmannian coordinates $\theta_\alpha$\,.  Expanding with respect to $\theta$\,, the superfield $Y^\mu$ of the superspace coordinates $(t\,, x^i, \theta_\alpha)$ takes the following form: 
\be
	Y^\mu (t\,, x^i, \theta_\alpha) = X^\mu (t, x^i) + \overline{\theta} \, \Psi^\mu (t, x^i) + \tfrac{1}{2} \, \overline{\theta} \theta \, B^\mu (t, x^i).
\ee
The fermionic transformations in \eqref{eq:susytrsf} can be written compactly as
\be \label{eq:susytrnsfssp}
	\delta_\epsilon Y^\mu = \bigl[ \overline{\epsilon} \, \CQ\,, Y^\mu \bigr]\,,
		\qquad
	\CQ_\alpha \equiv \frac{\p}{\p {\overline{\theta}}^\alpha} - (\rho \, \theta)_\alpha \, \p_t\,. 
\ee
In this operator representation, $P_0 = - i \, \p_t$\,. Define the supercovariant derivative 
\be \label{eq:Dalpha}
	D_\alpha = \frac{\p}{\p {\overline{\theta}}^\alpha} + ( \rho \, \theta )_\alpha \, \p_t\,,
\ee
such that $\{ \CQ\,, D\} = 0$\,. The action that contains the most general marginal terms invariant under the transformation \eqref{eq:susytrnsfssp} is \cite{ssl, as}\footnote{We use the convention $\int d^2 \theta \, \overline{\theta} \theta = 1$\,. At the RG fixed point $G_{\mu\nu} = H_{\mu\nu}$\,, the theory is closely related to the effective action in the context of stochastic quantization with the Parisi-Sourlas supersymmetry (see, e.g., \cite{ZinnJustin:1986eq}). At equilibrium, the action \eqref{eq:tSbrane} in the single metric limit reduces to a two-dimensional Euclidean theory, which coincides with the action of the harmonic topological sigma model on a flat worldsheet \cite{Horava:1993aq, Horava:1995ic}.}
\begin{align} \label{eq:tSbrane}
	\tilde{S}_\text{brane} = \frac{\tilde{T}_2}{2} \int dt \, d^2 \mathbf{x} \, d^2 \theta \, \Bigl\{ \overline{D}_\alpha Y^\mu \, D_\alpha Y^\nu \, G_{\mu\nu} (Y) - 2 \, \p_i Y^\mu \, \p_i Y^\nu \, H_{\mu\nu} (Y) \Bigr\}\,. 
\end{align}
Here, the superfield $Y$ is dimensionless and the theory is at its lower critical dimensions. This sigma model is power-counting renormalizable. Coupling the sigma model \eqref{eq:tSbrane} to worldvolume gravity requires supersymmetrizing $(2+1)$-dimensional Weyl-invariant Ho\v{r}ava gravity, for which techniques developed in \cite{Frenkel:2020djn, Frenkel:2020ixs, Frenkel:2020dic} can be borrowed.

In the case when the deformation parametrized by the speed of light $c$ in the anti-commutation relation \eqref{eq:QQc} is turned on, an associated deformation in \eqref{eq:Dalpha} is also generated,
\be \label{eq:Dalphac}
	D_\alpha = \frac{\p}{\p \overline{\theta}^\alpha} + (\rho \, \theta)_\alpha \, \p_t + c \, (\rho^i \theta)_\alpha \, \p_i\,.
\ee
Substituting \eqref{eq:Dalphac} in \eqref{eq:tSbrane} gives rise to the NLSM that flows towards three-dimensional relativistic $\CN=1$ supersymmetric NLSM in the IR.

Integrating out $\theta$ and the auxiliary field $B^\mu$ in \eqref{eq:tSbrane}, we find that the bosonic part of the action is
\be \label{eq:bosonicpart}
	\tilde{S}^{\text{B}}_\text{brane} = \frac{\tilde{T}_2}{2} \int dt \, d^2 \mathbf{x} \, \Bigl( \p_t X^\mu \, G_{\mu\nu} \, \p_t X^\nu - \Delta^{\!i} \p_i X^\mu \, H_{\mu\rho} \, G^{\rho\sigma}  H_{\sigma\nu} \, \Delta^{\!j} \p_j X^\nu \Bigr)\,.
\ee
This significantly simplifies the bosonic action \eqref{eq:Sbrane} while retaining the bimetric natural of the target space. It is manifest that the matter fields have a quadratic dispersion relation. 

The supersymmetric NLSM describes classical membranes propagating in a bimetric spacetime. However, unlike the two-dimensional sigma model \eqref{eq:action2} in a bimetric spacetime, which is self-dual under the transformation \eqref{eq:swap}, now, in \eqref{eq:tSbrane}, the metric fields $G_{\mu\nu}$ and $H_{\mu\nu}$ are \emph{not} on the same footing anymore. Therefore, we expect genuinely distinct beta-functionals for $G_{\mu\nu}$ and $H_{\mu\nu}$ in the membrane action \eqref{eq:tSbrane}. Power-counting renormalizable NLSMs of this type can also be constructed in higher dimensions. These NLSMs exhibit different Lifshitz scaling symmetries and in principle lead to an infinite hierarchy of multi-tensorial target-space geometries.

\section{Conclusions} \label{sec:concl}

In this paper, we considered a novel type of two-dimensional NLSMs, defined on a nonrelativistic worldsheet that lacks any local (Lorentzian nor Galilean) boost symmetries. The worldsheet dynamics is described by topological Ho\v{r}ava gravity. Imposing the worldsheet time-reversal symmetry, the sigma model is coupled to a pair of metric fields and describes classical strings propagating in a bimetric spacetime. We analyzed the RG flows in the bimetric sigma model, and derived the beta-functionals of the bimetric fields in \eqref{eq:betaGH}, up to the fourth-order in the Taylor expansion with respect to a small deviation away from equal metric. This set of beta-functionals form a pair of coupled Ricci flow equations. 
In the limit where the two metric fields are identical to each other, the coupled Ricci flow equations reduce to a single Ricci flow equation that arises as the beta-functional in relativistic string theory. Imposing Weyl invariance at the quantum level, we discover the equations of motion that govern the dynamics of the background geometry in the bimetric sigma model. In the linearized bimetric gravity, there emerges a massless gravitational excitation, which is accompanied by other degrees of freedom that satisfy the Klein-Gordon equation in \eqref{eq:leom}. 

Our study of Lifshitz-type sigma models hints at new bimetric gravity theories. Different versions of bimetric and multimetric gravity theories have been studied extensively in the literature. A general formulation of bimetric gravities is introduced in \cite{Rosen:1975kk} (see also \cite{Rosen:1940zza, Rosen:1940zz}), using which a special version of bimetric gravity is put forward and shown to be free of singularities while preserving the general covariance principles \cite{Rosen:1980dp}. Along other lines, in the context of massive gravity, a modern version of bimetric gravity (see, e.g., \cite{Hassan:2011zd, Bergshoeff:2013xma} and references therein) has been constructed, which has the attractive feature of being free of the Boulware-Deser ghosts \cite{Boulware:1973my} and also bears applications to cosmology,
the electroweak hierarchy problem \cite{Avgoustidis:2020wrd}, the fractional quantum Hall effect in condensed matter theory \cite{Gromov:2017qeb}, \emph{et cetera}. Furthermore, it has been shown in \cite{Kiritsis:2008at} that related multi-metric description of massive gravity also arises in the context of non-local multi-string theory \cite{Aharony:2001pa}.
Nevertheless, the bimetric gravity considered in our work does not seem to fit into any of the previously existing models. To further the understanding of the bimetric gravity that arises in our work, and to identify its relation to existing bimetric theories, one imminent future task is to construct a gravitational action principle that gives rise to the spacetime equations of motion found in this paper and analyze the Hamiltonian constraints to determine the actual degrees of freedom. The fact that our bimetric gravity arises from a string-theoretical setup makes it promising that the full interacting theory might be free of ghosts, under the condition that the worldsheet theory can be shown to be unitary.

The main focus of this paper has been the beta-functionals of the spacetime bimetric fields. However, to map out the complete RG structure of the two-dimensional Lifshitz-type sigma model, one also needs to determine the beta-functionals for the dilaton fields, for which detailed RG properties of the worldsheet Ho\v{r}ava gravity is required (some useful techniques have been developed in the literature, see, e.g., \cite{DOdorico:2015pil, Barvinsky:2015kil, Griffin:2017wvh, Barvinsky:2017mal, Barvinsky:2017kob, Grosvenor:2021zvq}). This piece of calculation will be essential for understanding the notion of critical dimensions in Lifshitz-type sigma models. Further investigations of the foliated worldsheet topology will be important for revealing whether there exists a well-defined perturbative expansion with respect to a unique effective string coupling, formed by the two dilaton fields in \eqref{eq:total}. It will be fascinating to find out whether our nonrelativistic sigma model can ultimately be promoted to describe a self-consistent quantum theory of strings that generalizes relativistic string theory.

The concepts and techniques developed in this paper are applicable to a large class of Lifshitz-type sigma models that map $p$-branes to novel spacetime geometries described by multi-tensorial fields. We have discussed two natural generalizations in the paper. The first variant that we introduced in \S\ref{sec:trb} is by explicitly breaking the time-reversal symmetry in our sigma model, which leads to a two-dimensional sigma model in a trimetric spacetime geometry, coupled to a $B$-field and multiple dilaton fields. Later in \S\ref{sec:omqc}, we studied a second variant that is a three-dimensional supersymmetric sigma model describing membranes at quantum criticality, following \cite{ssl, as}. Understanding these new ingredients will boost the exploration of the mostly uncharted territory of Lifshitz-type NLSMs, providing an arena for probing new geometries and alternative constructions of quantum membranes. 

\acknowledgments

The author is grateful to Petr Ho\v{r}ava and Charles Melby-Thompson for discussions and collaborations that inspired this work. It is also a pleasure to thank Shira Chapman, Kevin T. Grosvenor, Florian Niedermann, Niels A. Obers, Jan Rosseel, and Igal Arav for useful discussions and comments on a draft of this paper. Thanks to F. Niedermann and J. Rosseel for useful insights on massive gravity. 

\newpage

\appendix

\section{Exact Heat Kernel Coefficient} \label{app:chkc}

In \eqref{eq:sigma2result}, we presented the terms in $\sigma_2$ that contribute nontrivially the heat kernel coefficient $E_2$\,. 
We then performed a perturbative expansion with respect to $\epsilon$ introduced in \eqref{eq:GminusH} and kept terms in $E_2$ up to the fourth order in $\epsilon$\,. In this appendix, we take a step forward and compute $E_2$ exactly to all orders in $\epsilon$\,. 
We will set $\epsilon = 1$ in the following discussion and the Taylor expansions will be taken with respect to $f$ directly. 
We start with transcribing the result for $\sigma_2^{\mu\nu}$ in \eqref{eq:sigma2result} (where only terms that make nonzero contributions to $E_2$ are kept):
\be
	[\sigma_2^{\mu\nu}] \sim \sum_{a=1}^{10} I_a^{\mu\nu} + I_V\,,  
\ee
where
\begin{subequations}
\begin{align}
	I_1^{\mu\nu} & = - 4 \, \omega^2 \, \kappa^4 \, \CD^{\mu\rho} \, \CD^{\sigma\lambda} \, \CD^{\kappa\theta} \, \CD^{\xi\zeta}  \, \CD^{\eta\nu} \, G_{\rho\sigma} \, \nabla_{\!\tau} H_{\lambda\kappa} \, G_{\theta\xi} \, \nabla_{\!\tau} H_{\zeta\eta}\,, \\[2pt]
	I_2^{\mu\nu} & = - 8 \, \omega^2 \, \kappa^4 \, \CD^{\mu\rho} \, \CD^{\sigma\lambda} \, \CD^{\kappa\theta} \, \CD^{\xi\zeta}  \, \CD^{\eta\nu} \, G_{\rho\sigma} \, G_{\lambda\kappa} \, \nabla_{\!\tau} H_{\theta\xi} \, \nabla_{\!\tau} H_{\zeta\eta}\,, \\[2pt]
	I_3^{\mu\nu} & = 4 \, \omega^2 \, \kappa^2 \, \CD^{\mu\rho} \, \CD^{\sigma\lambda} \, \CD^{\kappa\theta} \, \CD^{\xi\nu} \, G_{\rho\sigma} \, G_{\lambda\kappa} \,  \nabla^2_{\!\tau} H_{\theta\xi}\,, \\[2pt]
	I_4^{\mu\nu} & = - 4 \, \kappa^2 \, \Omega^4 \, \CD^{\mu\rho} \, \CD^{\sigma\lambda} \, \CD^{\kappa\theta} \, \CD^{\xi\zeta}  \, \CD^{\eta\nu} \, H_{\rho\sigma} \, H_{\theta\xi} \, \Delta_x G_{\lambda\kappa} \, \Delta_x G_{\zeta\eta}\,, \\[2pt]
	I_5^{\mu\nu} & = - 8 \, \kappa^2 \, \Omega^4 \, \CD^{\mu\rho} \, \CD^{\sigma\lambda} \, \CD^{\kappa\theta} \, \CD^{\xi\zeta}  \, \CD^{\eta\nu} \, H_{\rho\sigma} \, H_{\lambda\kappa} \, \Delta_x G_{\theta\xi} \, \Delta_x G_{\zeta\eta}\,, \\[2pt]
	I_6^{\mu\nu} & = 4 \, \kappa^2 \, \Omega^2 \, \CD^{\mu\rho} \, \CD^{\sigma\lambda} \, \CD^{\kappa\theta} \, \CD^{\xi\nu} \, H_{\rho\sigma} \, H_{\lambda\kappa} \,  \Delta^2_{x} G_{\theta\xi}\,, \\[2pt]
	I_7^{\mu\nu} & = 2 \, \kappa^4 \, \CD^{\mu\rho} \, \CD^{\sigma\lambda} \, \CD^{\kappa\theta} \, \CD^{\xi\nu} \, G_{\rho\sigma} \, \nabla_{\!\tau} H_{\lambda\kappa} \,  \nabla_{\!\tau} H_{\theta\xi}\,, \\[2pt]
	I_8^{\mu\nu} & = - \kappa^2 \, \CD^{\mu\rho} \, \CD^{\sigma\lambda} \, \CD^{\kappa\nu} \, G_{\rho\sigma} \, \nabla^2_{\!\tau} H_{\lambda\kappa}\,, \\[2pt]
	I_9^{\mu\nu} & = 2 \, \Omega^4 \, \CD^{\mu\rho} \, \CD^{\sigma\lambda} \, \CD^{\kappa\theta} \, \CD^{\xi\nu} \, H_{\rho\sigma} \, \Delta_{x} G_{\lambda\kappa} \, \Delta_{x} G_{\theta\xi}\,, \\[2pt]
	I_{10}^{\mu\nu} & = - \Omega^2 \, \CD^{\mu\rho} \, \CD^{\sigma\lambda} \, \CD^{\kappa\nu} \, H_{\rho\sigma} \, \Delta^2_{x} G_{\lambda\kappa}\,,
\end{align}
\end{subequations}
and $I_V$ has been defined in \eqref{eq:IV}. The heat kernel coefficient $E_2$ is then given by
\be \label{eq:E2sum}
	E_2 = \sum_{a=1}^{10} \CI_a + \CI_V\,,
\ee
where $\CI_V$ has been computed in \eqref{eq:CIV} and
\be
	\CI_a = \int \frac{d\omega \, d\kappa}{(2\pi)^2} \int_\CC \frac{i \, d\lambda}{2\pi} \, e^{-\lambda} \, G_{\mu\nu} \, I_a^{\mu\nu}\,,
		\qquad
	a = 1, \cdots, 10\,.
\ee
The results for $\CI_a$ are computed below:
\begin{align*}
	\CI_1 & = - \frac{1}{2\pi^{3/2}} \, \nabla_{\!\tau} H_{\mu\nu} \, \nabla_{\!\tau} H_{\rho\sigma} 
		\!\!\!\!\!\! \sum_{n_1,\, \cdots,\, n_5 = 0}^\infty \frac{\Gamma\bigl( \frac{5}{2} + \sum_{a=1}^5 n_a\bigr)}{\Gamma \bigl( 5 + \sum_{a=1}^5 n_a \bigr)} \, \bigl( g^{n_1 + n_2 + n_5} \, G^{-1} \bigr)^{\mu\rho} \, \bigl( g^{n_3 + n_4} \, G^{-1} \bigr)^{\nu\sigma} \notag \\[2pt]
		& = - \frac{1}{64 \pi} \, G^{\mu\rho} \, G^{\nu\sigma} \, \nabla_{\!\tau} f_{\mu\nu} \nabla_{\!\tau} f_{\rho\sigma} - \frac{5}{128 \pi} \, G^{\rho\sigma} \, G^{\lambda\nu} \, g^\mu{}_\lambda \nabla_{\!\tau} f_{\mu\rho} \nabla_{\!\tau} f_{\nu\sigma} \notag \\[4pt]
		& \quad\hspace{0.5mm} - \frac{7}{512 \pi} \, \bigl( 2 \, G^{\nu\lambda} \, G^{\sigma \kappa} \, g^\mu{}_\lambda \, g^\rho{}_\kappa \nabla_{\!\tau} f_{\mu\rho} \nabla_{\!\tau} f_{\nu\sigma} + 3 \, G^{\rho\sigma} \, G^{\lambda \kappa} \, g^\mu{}_\lambda \, g^\nu{}_\kappa \nabla_{\!\tau} f_{\mu\rho} \nabla_{\!\tau} f_{\nu\sigma} \bigr) + O (f^5)\,, \\[10pt] 
	\CI_2 & = - \frac{1}{\pi^{3/2}} \, \nabla_{\!\tau} H_{\mu\nu} \, \nabla_{\!\tau} H_{\rho\sigma} 
		\!\!\!\!\!\! \sum_{n_1, \, \cdots, \, n_5 = 0}^\infty \frac{\Gamma \bigl( \tfrac{5}{2} + \sum_{a = 1}^5 n_a \bigr)}{\Gamma \bigl( 5 + \sum_{a = 1}^5 n_a \bigr)} \, \bigl( g^{n_1 + n_2 + n_3 + n_5} \, G^{-1} \bigr)^{\mu\rho} \, \bigl( g^{n_4} \, G^{-1} \bigr)^{\nu\sigma} \notag \\[2pt]
		& = - \frac{1}{32\pi} \, G^{\mu\rho} \, G^{\nu\sigma} \, \nabla_{\!\tau} f_{\mu\nu} \nabla_{\!\tau} f_{\rho\sigma} - \frac{5}{64 \pi} \, G^{\rho\sigma} \, G^{\lambda\nu} \, g^\mu{}_\lambda \nabla_{\!\tau} f_{\mu\rho} \nabla_{\!\tau} f_{\nu\sigma} \notag \\[4pt]
		& \quad\hspace{0.5mm} - \frac{7}{768 \pi} \, \bigl( 4 \, G^{\nu\lambda} \, G^{\sigma \kappa} \, g^\mu{}_\lambda \, g^\rho{}_\kappa \nabla_{\!\tau} f_{\mu\rho} \nabla_{\!\tau} f_{\nu\sigma} + 11 \, G^{\rho\sigma} \, G^{\lambda \kappa} \, g^\mu{}_\lambda \, g^\nu{}_\kappa \nabla_{\!\tau} f_{\mu\rho} \nabla_{\!\tau} f_{\nu\sigma} \bigr) + O (f^5)\,, \\[10pt]
	\CI_3 & = \frac{1}{24 \pi} \, \nabla_{\!\tau}^2 H_{\mu\nu} \, \bigl[ \bigl( G^{-1} H \bigr)^{-3/2} \, G^{-1} \bigr]^{\mu\nu} \notag \\[2pt]
		& = - \frac{1}{24\pi} \, G^{\mu\nu} \, \nabla_{\!\tau}^2 f_{\mu\nu} - \frac{1}{16\pi} \, G^{\mu\rho} \, G^{\nu\sigma} \, f_{\mu\nu} \, \nabla_{\!\tau}^2 f_{\rho\sigma} - \frac{5}{64\pi} \, G^{\rho\sigma} \, g^\mu{}_\rho \, g^\nu{}_\sigma \, \nabla_{\!\tau}^2 f_{\mu\nu} \notag \\[4pt]
		& \quad\hspace{0.5mm} - \frac{35}{384 \pi} \, G^{\lambda\nu} \, g^\mu{}_\lambda \, g^\rho{}_\mu \, g^\sigma{}_\nu \nabla_{\!\tau}^2 f_{\rho\sigma} + O(f^5) \,, \\[10pt]
	\CI_4 & = - \frac{15}{4\pi^{3/2}} \, \Delta_x G_{\mu\nu} \, \Delta_x G_{\rho\sigma} \notag \\[2pt]
		& \qquad\quad \times \!\!\!\!\!\! \sum_{n_1, \, \cdots, \, n_5 = 0}^\infty \frac{\Gamma \bigl( \tfrac{3}{2} + \sum_{a = 1}^5 n_a \bigr)}{\Gamma \bigl( 5 + \sum_{a = 1}^5 n_a \bigr)} \, \bigl( G^{-1} H \, g^{n_1 + n_2 + n_5} \, G^{-1} \bigr)^{\mu\rho} \, \bigl( G^{-1} H \, g^{n_3 + n_4} \, G^{-1} \bigr)^{\nu\sigma} \notag \\[4pt]
		& = - \frac{5}{64\pi} \, G^{\mu\rho} \, G^{\nu\sigma} \, \Delta_{x} f_{\mu\nu} \, \Delta_{x} f_{\rho\sigma} + \frac{5}{128 \pi} \, G^{\rho\sigma} \, G^{\lambda\nu} \, g^\mu{}_\lambda \, \Delta_{x} f_{\mu\rho} \, \Delta_{x} f_{\nu\sigma} \notag \\[4pt]
		& \quad\hspace{0.5mm} - \frac{5}{512 \pi} \, \bigl( 2 \, G^{\nu\lambda} \, G^{\sigma \kappa} \, g^\mu{}_\lambda \, g^\rho{}_\kappa \, \Delta_{x} f_{\mu\rho} \, \Delta_{x} f_{\nu\sigma} - 3 \, G^{\rho\sigma} \, G^{\lambda \kappa} \, g^\mu{}_\lambda \, g^\nu{}_\kappa \, \Delta_{x} f_{\mu\rho} \, \Delta_{x} f_{\nu\sigma} \bigr) + O (f^5) \,, \\[10pt]
	\CI_5 & = - \frac{15}{2\pi^{3/2}} \, \Delta_x G_{\mu\nu} \, \Delta_x G_{\rho\sigma} \notag \\[2pt]
		& \qquad\quad \times \!\!\!\!\!\! \sum_{n_1, \, \cdots, \, n_5 = 0}^\infty \frac{\Gamma \bigl( \tfrac{3}{2} + \sum_{a = 1}^5 n_a \bigr)}{\Gamma \bigl( 5 + \sum_{a = 1}^5 n_a \bigr)} \, \bigl[ (G^{-1} H)^2 \, g^{n_1 + n_2 + n_3 + n_5} \, G^{-1} \bigr]^{\mu\rho} \, \bigl( g^{n_4} \, G^{-1} \bigr)^{\nu\sigma} \notag \\[4pt]
		& = - \frac{5}{32\pi} \, G^{\mu\rho} \, G^{\nu\sigma} \, \Delta_{x} f_{\mu\nu} \, \Delta_{x} f_{\rho\sigma} + \frac{5}{64 \pi} \, G^{\rho\sigma} \, G^{\lambda\nu} \, g^\mu{}_\lambda \, \Delta_{x} f_{\mu\rho} \, \Delta_{x} f_{\nu\sigma} \notag \\[4pt]
		& \quad\hspace{0.5mm} + \frac{1}{256 \pi} \, \bigl( 4 \, G^{\nu\lambda} \, G^{\sigma \kappa} \, g^\mu{}_\lambda \, g^\rho{}_\kappa \, \Delta_{x} f_{\mu\rho} \, \Delta_{x} f_{\nu\sigma} + G^{\rho\sigma} \, G^{\lambda \kappa} \, g^\mu{}_\lambda \, g^\nu{}_\kappa \, \Delta_{x} f_{\mu\rho} \, \Delta_{x} f_{\nu\sigma} \bigr) + O (f^5) \,, \\[10pt]
	\CI_6 & = \frac{1}{8 \pi} \, \Delta_{x}^2 G_{\mu\nu} \, \Bigl[ \bigl( G^{-1} H \bigr)^{1/2} \, G^{-1} \Bigr]^{\mu\nu} \notag \\[2pt]
		& = \frac{1}{8\pi} \, G^{\mu\nu} \Delta_x^2 f_{\mu\nu} - \frac{1}{16\pi} \, G^{\mu\rho} \, G^{\nu\sigma} \, f_{\mu\nu} \, \Delta_{x}^2 f_{\rho\sigma} - \frac{1}{64\pi} \, G^{\rho\sigma} \, g^\mu{}_\rho \, g^\nu{}_\sigma \, \Delta_{x}^2 f_{\mu\nu} \notag \\[4pt]
		& \quad\hspace{0.5mm} - \frac{1}{128 \pi} \, G^{\lambda\nu} g^\mu{}_\lambda \, g^\rho{}_\mu \, g^\sigma{}_\nu \, \Delta_{x}^2 f_{\rho\sigma} + O(f^5) \,, \\[10pt]
		%
	\CI_7 & = \frac{1}{2\pi^{3/2}} \, \nabla_{\!\tau} H_{\mu\nu} \, \nabla_{\!\tau} H_{\rho\sigma}
		\!\!\!\!\!\! \sum_{n_1,\, \cdots,\, n_4 = 0}^\infty \frac{\Gamma\bigl( \frac{5}{2} + \sum_{a = 1}^4 n_a \bigr)}{\Gamma \bigl( 4 + \sum_{a = 1}^4 n_a \bigr)} \, \bigl( g^{n_1 + n_2 + n_4} \, G^{-1} \bigr)^{\mu\rho} \, \bigl( g^{n_3} \, G^{-1} \bigr)^{\nu\sigma} \notag \\[2pt]
		& = \frac{1}{16 \pi} \, G^{\mu\rho} \, G^{\nu\sigma} \, \nabla_{\!\tau} f_{\mu\nu} \nabla_{\!\tau} f_{\rho\sigma} + \frac{5}{32 \pi} \, G^{\rho\sigma} \, G^{\lambda\nu} \, g^\mu{}_\lambda \nabla_{\!\tau} f_{\mu\rho} \nabla_{\!\tau} f_{\nu\sigma} \notag \\[4pt]
		& \quad\hspace{0.5mm} + \frac{7}{256 \pi} \, \bigl( 3 \, G^{\nu\lambda} \, G^{\sigma \kappa} \, g^\mu{}_\lambda \, g^\rho{}_\kappa \nabla_{\!\tau} f_{\mu\rho} \nabla_{\!\tau} f_{\nu\sigma} + 7 \, G^{\rho\sigma} \, G^{\lambda \kappa} \, g^\mu{}_\lambda \, g^\nu{}_\kappa \nabla_{\!\tau} f_{\mu\rho} \nabla_{\!\tau} f_{\nu\sigma} \bigr) + O (f^5)\,, \\[10pt] 
	\CI_8 & = - \frac{1}{16\pi} \, \nabla_{\!\tau}^2 H_{\mu\nu} \, \bigl[ \bigl( G^{-1} H \bigr)^{-3/2} \, G^{-1} \bigr]^{\mu\nu} = - \frac{3}{2} \, \CI_3\,, \\[10pt]
	\CI_9 & = \frac{15}{8 \pi^{3/2}} \, \Delta_x G_{\mu\nu} \, \Delta_x G_{\rho\sigma}
		\!\!\!\!\!\! \sum_{n_1,\, \cdots,\, n_4 = 0}^\infty \frac{\Gamma\bigl( \frac{1}{2} + \sum_{a = 1}^4 n_a \bigr)}{\Gamma \bigl( 4 + \sum_{a = 1}^4 n_a \bigr)} \, \bigl( G^{-1} H \, g^{n_1 + n_2 + n_4} \, G^{-1} \bigr)^{\mu\rho} \, \bigl( g^{n_3} \, G^{-1} \bigr)^{\nu\sigma} \notag \\[2pt]
		& = \frac{5}{16\pi} \, G^{\mu\rho} \, G^{\nu\sigma} \, \Delta_{x} f_{\mu\nu} \, \Delta_{x} f_{\rho\sigma} - \frac{5}{32 \pi} \, G^{\rho\sigma} \, G^{\lambda\nu} \, g^\mu{}_\lambda \, \Delta_{x} f_{\mu\rho} \, \Delta_{x} f_{\nu\sigma} \notag \\[4pt]
		& \quad\hspace{0.5mm} - \frac{1}{256 \pi} \, \bigl( G^{\nu\lambda} \, G^{\sigma \kappa} \, g^\mu{}_\lambda \, g^\rho{}_\kappa \, \Delta_{x} f_{\mu\rho} \, \Delta_{x} f_{\nu\sigma} + 9 \, G^{\rho\sigma} \, G^{\lambda \kappa} \, g^\mu{}_\lambda \, g^\nu{}_\kappa \, \Delta_{x} f_{\mu\rho} \, \Delta_{x} f_{\nu\sigma} \bigr) + O (f^5)\,, \\[10pt]
	\CI_{10} & = - \frac{3}{32 \pi} \, \Delta_{x}^2 G_{\mu\nu} \, \bigl[ \bigl( G^{-1} H \bigr)^{1/2} \, G^{-1} \bigr]^{\mu\nu} = - \frac{3}{2} \, \CI_6\,. 
\end{align*}
For comparison, we expanded $\CI_a$ up to the fourth order in $f$ after giving their exact expressions, written as infinite sums. Plugging the above expressions into \eqref{eq:E2sum} and keeping up to the fourth order in $f$ reproduces \eqref{eq:E2e4}.

\newpage

\bibliographystyle{JHEP}
\bibliography{smbs}

\end{document}